\documentclass[12pt]{article}

\usepackage{bbm}
\usepackage{epsfig}
\usepackage{array}
\usepackage{float}
\usepackage{dsfont}
\usepackage{amstext}
\usepackage{rotating}
\usepackage{a4}
\usepackage{a4wide}
\usepackage{cite}

\newcommand{\be}{\begin{eqnarray}}
\newcommand{\ee}{\end{eqnarray}}

\def\nue{{\nu_e}}

\def\numu{{\nu_{\mu}}}

\def\nutau{{\nu_{\tau}}}

\def\lsim{\:\raisebox{-0.5ex}{$\stackrel{\textstyle<}{\sim}$}\:}

\newcommand{\ms}{\Delta m^2_{21}}
\newcommand{\ma}{\Delta m^2_{31}}

\newcommand{\sss}{\sin^2 \theta_{12}}
\newcommand{\sch}{\sin^2 \theta_{13}}

\def\ltap{\ \raisebox{-.4ex}{\rlap{$\sim$}} \raisebox{.4ex}{$<$}\ }

\begin{document}

\title{
\hfill {\small \texttt{HRI-P-08-06-003}} 
\vskip 0.4cm
\Large \bf Lepton Masses 
in a Minimal Model 
with Triplet Higgs Bosons and 
$S_3$ Flavor Symmetry}
\author{
Manimala Mitra\thanks{email: \tt mmitra@hri.res.in},~~~
Sandhya Choubey\thanks{email: \tt sandhya@hri.res.in}
\\\\
{\normalsize \it$^a$Harish--Chandra Research Institute,}\\
{\normalsize \it Chhatnag Road, Jhunsi, 211019 Allahabad, India }\\ \\ 
}
\date{ \today}
\maketitle
\thispagestyle{empty}
\vspace{-0.8cm}
\begin{abstract}
\noindent  
Viable neutrino and charged lepton masses and mixings are 
obtained by imposing a $S_3 \times Z_4 \times Z_3$ flavor 
symmetry in a model with a few additional Higgs. 
We use two $SU(2)_L$ triplet Higgs which are 
arranged as a doublet of $S_3$, and 
standard model singlet  Higgs  which are also put as doublets of 
$S_3$. We break the $S_3$ symmetry in this minimal model by 
giving vacuum expectation values (VEV) to the additional 
Higgs fields. Dictated by the minimum condition for the  
scalar potential, we obtain certain VEV alignments which 
allow us to maintain $\mu-\tau$ symmetry in the 
neutrino sector, while breaking it maximally for the charged leptons. 
This helps us to simultaneously explain the hierarchical charged 
lepton masses, and the neutrino masses and mixings. 
In particular, we obtain maximal $\theta_{23}$ and zero $\theta_{13}$.
We allow for a mild breaking of the $\mu-\tau$ symmetry for the 
neutrinos and study the phenomenology. We give predictions for 
$\theta_{13}$ and the CP violating Jarlskog invariant $J_{CP}$,  
as a function of the $\mu-\tau$ symmetry breaking parameter. 
We also discuss possible collider signatures and 
phenomenology associated with lepton flavor violating 
processes. 
\end{abstract}

\newpage

\section{Introduction}

Proof of neutrino masses and mixing from a series of 
outstanding experimental efforts, spanning many decades and 
using neutrinos from 
myriad types of natural 
as well as man-made sources, 
have opened a window to physics beyond the standard model 
of particle physics. Though 
there still remains a lot 
to be learnt about neutrino properties, 
a lot is already known \cite{review}. 
The two mass squared differences\footnote{We define 
$m_{ij}^2 = m_i^2 - m_j^2$.} 
$\Delta m_{21}^2$ and 
$\Delta m_{31}^2$, and 
the two mixing angles $\theta_{12}$ and $\theta_{23}$  
are now fairly well determined. The third mixing 
angle $\theta_{13}$, though still not determined, 
is known to be small. How small, is the question which 
will be answered in the next generation 
oscillation experiments. The current $3\sigma$ 
allowed ranges of the 
oscillation parameters are given as \cite{limits}
\be
7.1\times 10^{-5} eV^2 < \ms < 8.3\times 10^{-5} eV^2~,~~
2.0\times 10^{-3} eV^2 < \ma < 2.8\times 10^{-3} eV^2~,~~
\label{eq:dmconstraint}
\ee
\be
0.26<\sss<0.42~,~~\sin^22\theta_{23}>0.9~,~~\sch<0.05~.~~
\label{eq:thconstraint}
\ee
The best-fit of the global neutrino oscillation data
corresponds to 
maximal $\theta_{23}$  and zero $\theta_{13}$. 
This has prompted the speculation that a $\mu-\tau$ 
permutation symmetry \cite{mutau} might exist in the neutrino 
sector\footnote{A $L_\mu-L_\tau$ family symmetry 
could also very naturally give maximal $\theta_{23}$ 
and zero $\theta_{13}$ \cite{lmultau}.}. 
This symmetry should however be broken for the 
charged leptons, for which we know that a hierarchy 
exists between the masses of the $\mu$ and $\tau$. 
It cannot exist in the quark sector either, where the 
hierarchy between the masses is larger and mixing 
angles are known to be very small. In addition, 
the fact that the best-fit solar mixing angle is close 
to $\sss=1/3$ suggests that the lepton mixing matrix 
has the tribimaximal (TBM) mixing form 
\cite{tbm}.
Challenge for model builders lies in constructing an 
aesthetically simple and phenomenologically viable model, 
which could explain all aspects of the fermion masses and mixing. 
One way of generating the observed pattern of fermion masses 
is to impose certain flavor symmetries. The $A_4$ group 
has received a lot of attention recently \cite{a4,AF,a4us} as a 
way of generating TBM mixing for the neutrinos. However, there 
are some drawbacks of 
these models. The basic version of some of these models 
might need some fine tuning \cite{a4us}. 
They are also not able to give a consistent 
explanation of the quark mass hierarchies and the CKM mixing. 
Various extensions of the models with $A_4$ family symmetry have 
been proposed \cite{a4extension} in order to address these issues. 
However, most of these 
require additional discrete symmetries and many more scalar particles.

Another discrete group which has been extensively 
discussed in the literature as a family symmetry group 
is the $S_3$ permutation group \cite{s3old,s3hs,s3z2,s3tbm,s3DMT,s3,s3FL}. 
Many of these 
models predict TBM mixing. However, the main challenge 
still is to predict TBM mixing for neutrinos and at the same 
time reproduce an almost diagonal charged lepton mass matrix with 
the correct mass hierarchies. The quark sector also needs to be 
explained. 
Most models considered so far use right-handed neutrinos 
and type I seesaw for generating the neutrino masses. Only in 
\cite{s3tbm} the authors consider a model with triplet Higgs, and 
hence employ a type I+II seesaw 
mechanism to generate the correct neutrino masses and mixing. 
We will present a model which does not have any right-handed 
neutrinos and uses $SU(2)_L$ triplet Higgs to generate 
Majorana neutrino masses. This has never been considered before. 
We show how small neutrino masses can be explained in this model 
without invoking the seesaw mechanism. We also show how the neutrino 
mixing can be explained naturally by imposing the 
discrete flavor symmetry $S_3$.

The $S_3$ group has the $S_2$ permutation group as its subgroup. 
If one identifies this subgroup with the $\mu-\tau$ exchange 
symmetry, then it is straightforward to get vanishing 
$\theta_{13}$ and maximal $\theta_{23}$ for the neutrinos. 
However, since the same group acts on the charged leptons as well,
this would lead to 
$\mu$ and $\tau$ masses of the same order. In addition
this would lead to a highly non-diagonal 
mass matrix for the charged leptons, which is undesirable in this 
case. Therefore, the $S_3$ group should be broken in such a way 
that $\mu-\tau$ permutation symmetry remains intact for the 
neutrinos but gets badly broken for the charged leptons. 
In a recent paper \cite{s3FL}, the authors have used this 
idea to generate a viable scenario which explains almost all 
aspects of the lepton masses within a framework of a $S_3\times Z_3$
family symmetry. They also extend the model with additional 
$Z_3$ symmetries and Higgses to explain also the mass 
and mixing pattern of the quarks. In order to get the desired 
mass matrices, it is mandatory to have certain alignment for 
the vacuum expectation values (VEV) of the extra standard model 
singlet Higgs particles, which transform non-trivially under 
$S_3$. The authors of \cite{s3FL} use a supersymmetric version of their 
model with a few extra driving fields 
to explain the required VEV alignments. 

In this paper, we propose a model with $S_3\times Z_4\times Z_3$ family 
symmetry. The additional $Z_3$ symmetry is required 
for obtaining the correct form of the charged lepton mass matrix. 
As in \cite{s3FL},  
we preserve the $\mu-\tau$ symmetry in the neutrino sector while 
breaking it almost maximally for the charged leptons. 
Note that the field 
content and hence the mass generation mechanism 
of our model is completely different. In particular, 
we introduce two $SU(2)_L$ triplet Higgs in our model for 
generating the neutrino masses. The charged lepton masses are 
generated by the standard Higgs doublet. 
We postulate two additional $S_3$ doublets of Higgs which are 
$SU(2)_L\times U(1)_Y$ singlets, to 
generate  the desired lepton mass matrices.
The $S_3$ group 
is broken spontaneously when the 
singlet Higgs acquire VEVs. 
The VEVs are aligned in such a way that the residual 
$\mu-\tau$ symmetry is intact for the neutrinos but 
broken maximally for the charged leptons. 
We justify the VEV alignment by explicitly minimizing 
our scalar potential. We do not need to 
impose supersymmetry. We show that under the most general 
case, the minimization condition of our scalar potential 
predicts a very mild breaking of the $\mu-\tau$ 
symmetry for the neutrinos.  
We study the phenomenological viability
and testability of our model both in the exact as well as 
approximate $\mu-\tau$ symmetric cases. 
We give predictions for 
the mass squared differences, mixing angles,  
absolute neutrino mass scale, beta decay and 
neutrino-less double beta decay. 
It should be possible to extend our model to reproduce the quark mass 
hierarchy and CKM mixing\footnote{Work in progress.} such that the 
complete model is anomaly free\cite{anomaly}.

The paper is organized as follows: We give  an 
overview of the $S_3$ group in the appendix. In section 2
we introduce the particle content of our model and 
write down the mass matrices for the neutrinos and 
charged leptons. 
In section 3 we present the phenomenological 
implications of our model in the exact and approximate 
$\mu-\tau$ symmetric case. We discuss in detail 
the possible collider phenomenology and lepton 
flavor violating channels 
which could be 
used to provide smoking gun evidence for our model. 
Section 4 is devoted to justifying 
the alignment needed for the Higgs VEVs.
We end in section 
5 with our conclusions.

\section{The Model} 

\begin{table} 
\centering
\begin{tabular}{|c||c|c|c|c|c|c|c||c|c||c|}
\hline
Field & $H$ & $l_1$ & $D_l$ & $\overline{e_R}$ 
& $\overline{\mu_R}$ & $\overline{\tau_R}$ &  $\Delta$ & $\phi_e$  & $\xi$ \\
\hline
$S_3$ &1 & 1 & 2 & $1$ & $1$ & $1$ & 2 & 2 & 2 \\[3pt]
\hline
$Z_4$ & $i$ & $-1$ &1 & 1 & $-i$ & 1 & $-1$ & $i$ & $-1$ \\ [3pt]
\hline
$Z_3$ & 1 & 1 &1 & 1 & $\omega$ & $\omega^{2}$ & 1 & $\omega$ & 1\\
\hline
\end{tabular}
\caption{Transformation properties of matter and flavon fields 
under the flavor groups.}
\label{tab:charges3}
\end{table}

We present in Table \ref{tab:charges3} the particle 
content of our model and their transformation properties 
under the discrete groups $S_3$ , $Z_4$ and $Z_3$. The Higgs 
$H$ is the usual $SU(2)_L$ doublet, 
\be
H=\left(
\begin{array}{c}
{h^{+}}\\
{h^0}
\end{array}
\right)
~,~~~~
\label{eq:mch}
\ee 
which transforms as 
singlet under $S_3$. 
The Higgs $\Delta_1$ and $\Delta_2$ are $SU(2)_L$ triplets, 
\be
\Delta_i=\left(\begin{array}{cc}
\Delta_i^{+}/\sqrt{2}  & \Delta_i^{++}\\
\Delta_i^{0} & -\Delta_i^{+}/\sqrt{2} ~,
\end{array}
\right)~,
\ee
which transform as a doublet 
\be
\Delta
=\left(\begin{array}{c}
{\Delta_{1}}\\
{\Delta_{2}}
\end{array}
\right)~,
\ee 
under $S_3$. We introduce two additional 
$S_3$ scalar doublets $\phi_e$ and $\xi$,
\be
\phi_e=\left(\begin{array}{c}
{\phi_{1}}\\
{\phi_{2}}
\end{array}
\right)
,~~ 
\xi =\left(\begin{array}{c}
{\xi_{1}}\\
{\xi_{2}}
\end{array}
\right)~,
\ee
which are singlets 
under $SU(2)_L\times U_Y(1)$ and are hence our new flavon fields. 
The $SU(2)_L\times U_Y(1)$ lepton doublets are distributed in 
the $S_3$ multiplets as follows:
\be
D_l = \pmatrix{l_2 \cr l_3}~,
\ee
transforms as a doublet under $S_3$, where
$l_2 = (\numu_L, \mu_L)^T$ and $l_3 = (\nutau_L, \tau_L)^T$, 
while 
\be
l_1 = \pmatrix{\nue_L\cr e_L}~,
\ee
transforms as a singlet. The right-handed fields $e_R$, $\mu_R$ and 
$\tau_R$ transform as $1$ under $S_3$. The corresponding 
charges of the particles under $Z_4$ and $Z_3$ has been summarized in 
Table \ref{tab:charges3}. 

\subsection{Neutrino Masses and Mixing}

Given the field content of our model and their charge assignments 
presented in Table \ref{tab:charges3}, 
the most general $S_3\times Z_4\times Z_3$ invariant 
Yukawa part of the Lagrangian (leading order) giving the neutrino mass can 
be written as  
\be
 -{\cal L}^{y}_{\nu} = 
 \frac{y_2}{\Lambda}(D_{l}D_{l})^{\underline 1}
(\xi \Delta)^{\underline 1} + \frac{y_1}
{\Lambda}  (D_{l}D_{l})^{\underline 2}(\xi \Delta)^{\underline 2}
 + 2y_3 l_{1}D_{l}\Delta +  \frac{y_4}
{\Lambda} l_{1}l_{1}\xi \Delta+h.c.+...
\label{eq:nulag}
\ee
where $\Lambda$ is the cut-off scale of the theory and 
the underline sign in the superscript represents the 
particular $S_3$ representation from the tensor product of 
the two $S_{3}$ doublets\footnote{The term $(l_lD_l\Delta)$ denotes
$(l_l^TCi\tau_2D_l\Delta)$, where $C$ is the charge conjugation 
operator.}. 
Since $(D_lD_l)$ and $\xi\Delta$ are $2\times 2$ 
products which could give either $1$ or $2$, 
and since we can obtain $1$ either by $1\times 1$ or 
$2\times 2$, we have two terms coming from 
$(D_lD_l\xi\Delta)$. The  
$(D_lD_l)(\xi\Delta)$ as $1^\prime \times 1^\prime$ term 
does not contribute to the neutrino mass matrix.
Note that the presence of the $Z_4$ symmetry 
prevents the appearance of the usual 5 dimensional 
$D_lD_lHH$ and $l_1l_1HH$ Majorana mass term for the neutrinos. 
In fact, the neutrino mass matrix is completely 
independent of $H$ due to the $Z_4$ symmetry. In addition, 
there are no Yukawa couplings involving the neutrinos and 
the flavon $\phi_e$ due to  $Z_4$ or/and $Z_3$ symmetry. 
The $S_3$ symmetry is broken spontaneously when the 
flavon $\xi$ acquires a vacuum expectation value (VEV):  
\be
\langle \xi \rangle = \pmatrix{u_1 \cr u_2}
~.
\label{eq:xivev}
\ee

Finally, the $SU(2)_L\times U_Y(1)$ 
breaks at the electroweak scale giving VEVs to the triplets as
\be
\langle \Delta \rangle = \pmatrix{\langle \Delta_1 \rangle \cr 
\langle \Delta_2 \rangle},~~~~~{\rm where}~~
\langle \Delta_i \rangle = \pmatrix{0& 0 \cr v_i & 0}
~,
\label{eq:deltavev}
\ee
neutrinos get massive and their mass matrix is given as
\be
m_{\nu}=
 \left(
\begin{array}{ccc}
2y_{4}\frac{w}{\Lambda} & 2y_{3}v_{2} &  2y_{3}v_{1}\\
2y_{3}v_{2} & 2y_{1}\frac{u_{2}v_{2}}{\Lambda} & 2y_{2}\frac{w}{\Lambda} \\
2y_{3}v_{1} & 2y_{2}\frac{w}{\Lambda} &  2y_{1}\frac{u_{1}v_{1}}{\Lambda} 
\end{array}
\right)
~,
\label{eq:numassmatrix}
\ee
where $w = u_1v_2 + u_2v_1$. 
For the VEV alignments
\be
v_1 = v_2,~~{\rm and}~~u_1=u_2
~,
\label{eq:nuvev}
\ee
the neutrino mass matrix reduces to the form 
\be
m_{\nu}=
 \left(
\begin{array}{ccc}
2y_{4}\frac{2u_{1}v_{1}}{\Lambda} & 2y_{3}v_{1} &  2y_{3}v_{1}\\
2y_{3}v_{1} & 2y_{1}\frac{u_{1}v_{1}}{\Lambda} & 2y_{2}\frac{2u_{1}v_{1}}{\Lambda} \\
2y_{3}v_{1} & 2y_{2}\frac{2u_{1}v_{1}}{\Lambda} &  2y_{1}\frac{u_{1}v_{1}}{\Lambda}
\end{array}
\right)
~.
\label{eq:mnu}
\ee 
We will motivate our choice of the VEV alignments in 
section \ref{sec:scalar} where we will show that one can expect this from the minimization condition of the scalar potential. 
Denoting $ \frac{u_{1}} {\Lambda}$ as 
$u'_{1}$ the mass matrix becomes 
\be
m_{\nu}=
2v_{1} \left(
\begin{array}{ccc}
2y_{4}u'_{1} & y_{3} & y_{3}\\
y_{3} & y_1u'_{1} & 2y_{2}u'_{1}\\
y_{3} & 2y_{2}u'_{1} & y_1u'_{1}
\end{array}
\right)
~,
\ee
where $u'_{1}=\frac{u_{1}}{\Lambda}$ and it is less than 1.
Redefining  $2y_{4}u'_{1}$ as
$y_{4}$, $y_{1}u'_{1}$ as $y_{1}$ and $2y_{2}u'_{1}$ as $y_{2}$, 
the final form of the mass matrix is 
\be
m_{\nu}=
2v_{1} \left(
\begin{array}{ccc}
y_{4} & y_{3} & y_{3}\\
y_{3} & y_1 & y_{2}\\
y_{3} & y_{2} & y_1
\end{array}
\right)
~.
\label{eq:mnugen}
\ee
Note that if the VEV  $u_1=0$,
we would obtain the matrix
\be
m_{\nu}=
2v_{1}y_3 \left(
\begin{array}{ccc}
0 & 1 & 1\\
1 & 0 & 0\\
1 & 0 & 0
\end{array}
\right)
~.
\label{bimaxinverted}
\ee
This is a very well known form of the neutrino mass matrix. It 
returns inverted neutrino mass spectrum with eigenvalues 
$\{-2\sqrt{2}v_{1}y_3, 2\sqrt{2}v_{1}y_3,0\}$, and 
bimaximal mixing with $\theta_{23}=\theta_{12} = \pi/4$ and 
$\theta_{13}=0$. The only family symmetry considered in the 
literature for obtaining the form of the mass matrix 
given by Eq. (\ref{bimaxinverted}) is $L_e-L_\mu-L_\tau$ 
\cite{lelmlt}. We have here obtained this form of $m_\nu$ 
from a completely new kind of flavor symmetry. Of course 
exact bimaximal mixing is ruled out by the solar neutrino and 
KamLAND data \cite{limits}. Besides, as one can see from the 
eigenvalues of this neutrino mass matrix, that $\ms=0$. This is  
untenable in the light of the experimental data. In order to 
generate the correct $\ms$ and deviation of $\theta_{12}$ from 
maximal that is consistent with the data, 
one has to suitably perturb $m_\nu$ 
(for instance, see \cite{zeeus} as one example of such a 
model in the framework of the Zee-Wolfenstein ansatz). 
In the $S_3$ model that we consider here, this is very naturally 
obtained if we allow non-zero VEV for $\xi$. The strength of the 
additional terms is linearly proportional to $u_1/\Lambda$ and 
could be naturally small. 

In what follows, we will consider all values of $u_1/\Lambda$ from 
very small to $\sim 1$. 
The eigenvalues of the most general matrix given by 
Eq. (\ref{eq:mnugen})
are\footnote{For all analytical 
results given in this section we have assumed the model 
parameters to be real for simplicity. We check the phenomenological 
viability and testability of our model in the next section for complex 
Yukawas and VEVs.} 
\be 
m_{i}&=& v_{1}
\left(y_1 + y_{2} + y_{4} - \sqrt{y_1^{2}+
y^{2}_{2}+y^{2}_{4}+8y^{2}_{3}+2y_1y_{2}-2y_1y_{4}-2y_{2}y_{4}}\right) 
~,
\label{eq:mnueigen1}
\\ 
m_j&=& v_{1} \left(y_1 + y_{2} + y_{4} + \sqrt{y_1^{2}+
y^{2}_{2}+y^{2}_{4}+8y^{2}_{3}+2y_1y_{2}-2y_1y_{4}-2y_{2}y_{4}}\right)
~, 
\label{eq:mnueigen2}
\\
m_3&=&2v_{1}\bigg(y_1-y_{2}\bigg)
~.
\label{eq:mnueigen3}
\ee
Note that the only difference between $m_i$ and $m_j$ 
comes in the sign of the quantity within square root. 
We know that the solar neutrino data provides evidence for 
$\ms > 0$ at more than $6\sigma$ \cite{limits}. Therefore, 
the choice of $m_1$ and $m_2$ in Eqs. (\ref{eq:mnueigen1}) 
and (\ref{eq:mnueigen2}) 
is determined by the condition $m_2>m_1$ {\it viz.}, the 
larger eigenvalue corresponds to $m_2$. 
The eigenvectors are given as 
\be
U_{i} = \left(\begin{array}{c}
-\frac{y_1+y_{2}-y_{4} + \sqrt{a}}{2y_{3}b}\\
\frac{1}{b}\\
\frac{1}{b}
\end{array}
\right)
~,~~~~
U_j = 
\left(\begin{array}{c}
-\frac{y_1+y_{2}-y_{4}-\sqrt{a}}{2y_{3}c}\\
\frac{1}{c}\\
\frac{1}{c}
\end{array}
\right)
~,~~~~
U_3 = 
\left(\begin{array}{c}
0\\
-\frac{1}{\sqrt{2}}\\
\frac{1}{\sqrt{2}}\\
\end{array}
\right)
~,
\label{eq:mixing}
\ee
where $U_i$ corresponds to the eigenvalue given in 
Eq. (\ref{eq:mnueigen1}) and $U_j$ to that in 
Eq. (\ref{eq:mnueigen2}). Whether 
$U_1 \equiv U_i$ or $U_j$ depends on whether 
$m_i$ is smaller or larger than $m_j$. 
The quantities 
$b$ and $c$ are the normalization constants given 
by
\be
b^{2}=2+ \frac{(y_1+y_{2}-y_{4}+\sqrt{a})^{2}}{(2y_{3})^{2}} 
~,
\ee
and  
\be  
c^{2}=2+ \frac{(y_1+y_{2}-y_{4}-\sqrt{a})^{2}}{(2y_{3})^{2}} 
~,
\ee
and $a$ is given as
\be 
a=y_1^{2}+y^{2}_{2}+y^{2}_{4}+8y^{2}_{3}+2y_1y_{2}-2y_1y_{4}-
2y_{2}y_{4} 
~.
\ee  
From Eqs. (\ref{eq:mnueigen1}), (\ref{eq:mnueigen2}) and 
(\ref{eq:mnueigen3}) we obtain 
\be
\ms &=& 4\,v_1^2\,(y_1+y_2+y_4)\,\sqrt{a}
~,
\nonumber
\\
\ma & = & v_1^2\,(3y_1-y_2+y_4-\sqrt{a})(y_1-3y_2-y_4+\sqrt{a})
~.
\label{eq:massdiff}
\ee
The mixing angles can be seen from Eq. (\ref{eq:mixing}) to be 
\be
\theta_{13}^\nu&=&0
~,\nonumber
\\
\tan\theta_{23}^\nu& =& 1
~,\nonumber
\\
\tan\theta_{12}^\nu& =& \frac{(y_1+y_{2}-y_{4} - \sqrt{a})\,b}
{(y_1+y_{2}-y_{4}+\sqrt{a})\,c}
~.
\label{eq:mixnu}
\ee
Note that neither the ratio of the two mass squared differences
$\ms/\ma$, nor the mixing angles depend on the value of the 
triplet VEV $v_1$. They only depend on the 
Yukawas couplings. Only the absolute mass square 
differences $\ms$ and $\ma$ individually depend on the 
triplet VEV. 
The effective neutrino mass predicted for 
neutrino-less double beta decay is given as 
\be
|m_{\nu_{ee}}| = |2v_1y_4|
~,
\ee
while the effective mass squared observable in beta decay 
$m_\beta^2$ and 
the total neutrino mass crucial for cosmology $m_t$ are given as
\be
m_\beta^2 = \sum_i |m_i|^2 |U_{ei}|^2
~,~~{\rm and}~~~
m_t = \sum_i |m_i|
~,
\ee
respectively. 

\subsection{Charged Lepton Masses and Mixing}

The Yukawa Lagrangian up to order $1/\Lambda^3$ 
giving the charged lepton mass is 
\be
-{\cal L}^{y}_{e} &=& \frac{\gamma }
{\Lambda}\overline\tau_R H^\dagger(D_{l}\phi_{e})+
\frac{\gamma^{b}}
{\Lambda^3}\overline\tau_R H^\dagger(D_{l}\phi_{e})^{\underline 1}(\xi\xi)^{\underline 1}+
\frac{\gamma^{\prime}}
{\Lambda^3}\overline\tau_R H^\dagger(D_{l}\phi_{e})^{\underline
  2}(\xi\xi)^{\underline 2}
\nonumber \\
&&+\frac{\gamma^{\prime\prime\prime}}
{\Lambda^3}\overline\tau_R H^\dagger(D_{l}\phi'_{e})^{\underline
  1}{(\phi_{e}\phi_{e})}^{\underline 1}+
\frac{\gamma^{a}}
{\Lambda^3}\overline\tau_R H^\dagger(D_{l}\phi'_{e})^{\underline 2}{(\phi_{e}\phi_{e})}^{\underline 2}+
\frac{\gamma''}
{\Lambda^2}\overline\tau_R H^\dagger l_{1}(\phi_{e} \xi)
\nonumber \\ 
&&+
\frac{\beta'}
{\Lambda^2}\overline\mu_R H^\dagger (D_{l}\phi_{e}\phi_{e})+
\frac{\beta''}
{\Lambda^3}\overline\mu_R H^\dagger l_{1}(\phi_{e} \phi_{e} \xi)+
\frac{\alpha''}
{\Lambda^3}\overline e_R H^\dagger l_{1}(\phi_{e} \phi_{e} \phi_{e})\nonumber \\
&&+\frac{\alpha'}
{\Lambda^3}\overline e_R H^\dagger (D_{l}\phi'_{e})^{\underline 1}(\phi'_{e}\phi'_{e})^{\underline 1}+
\frac{\alpha}
{\Lambda^3}\overline e_R H^\dagger (D_{l}\phi'_{e})^{\underline 2}(\phi'_{e}\phi'_{e})^{\underline 2}+ h.c+... 
\label{eq:chlag}
\ee
While $Z_4$ symmetry was sufficient to get the desired $m_\nu$, 
the extra $Z_3$ symmetry had to be introduced in order to 
obtain the correct form for the charged lepton mass matrix. 
We reiterate that the presence of the $Z_4$ symmetry ensures that 
the flavon doublet $\phi_e$ couples to charged leptons only. 
This is a prerequisite since we wish to break $S_3$ 
such that the $\mu-\tau$ symmetry remains intact for 
the neutrinos while it gets maximally broken for the charged 
leptons. For neutrinos the 
$\mu-\tau$ symmetry was kept intact by the choice of 
the vacuum alignments given in Eq. (\ref{eq:nuvev}). To break 
it maximally for the charged leptons we choose the 
VEV alignment
\be
\langle \phi_e \rangle = \pmatrix{v_c\cr0}
~.
\label{eq:phivev}
\ee
Once $S_3$ is spontaneously broken 
by the VEVs of the flavons and $SU(2)_L\times U(1)_Y$ 
by the VEVs of the standard model doublet Higgs, 
we obtain the charged lepton 
mass matrix (leading terms only)\footnote{While this form of 
charged lepton mass matrix has been obtained using $Z_4 \times Z_3$
symmetry, similar viable forms can be obtained using other
$Z_n $ symmetries. For example, we have explicitly checked that 
$Z_6 \times Z_2$ and $Z_8 \times Z_2$ symmetries also give 
viable structure for $m_l$.}
\be
m_l = \pmatrix{\alpha^{\prime\prime} \lambda^2 & 0 & 0 \cr
\beta^{\prime\prime}\lambda u'_1 & \beta^{\prime}\lambda & 0 \cr
\gamma^{\prime\prime}u'_2 & \gamma^{\prime}{u'_2}^2 & \gamma \cr
}
v_{SM}\lambda
~,
\label{eq:ml}
\ee
where $v_{SM} = \langle H \rangle$ is  the VEV of the standard Higgs,
 $ \lambda = v_c/\Lambda$ , $u'_1= u_1/\Lambda$ and $u'_2= u_2/\Lambda$  
 The 
charged lepton masses and mixing matrix are obtained from 
\be
m_{l_{diag}}^2 = U_l\,m_lm_l^\dagger\,U_l^\dagger
~,
\ee
giving the masses as
\be
m_\tau \simeq \gamma \lambda v_{SM},~~~~
m_\mu \simeq   \beta^{\prime}\lambda^2 v_{SM},~~~~
m_e \simeq \alpha^{\prime\prime} \lambda^3 v_{SM}
~.
\label{eq:mleigen}
\ee
For  $\lambda \simeq 2\times 10^{-2}\simeq \frac{\lambda_c^2}{2}$ where 
$\lambda_c$ is the Cabibbo angle, the correct mass hierarchy of 
$\tau$ to $\mu$
to $e$ as well as their exact numerical values 
can be obtained by choosing 
 $\gamma=0.36$, $\beta'=1.01$ and 
$\alpha''=0.25$ and $v_{SM}=246$ GeV. Note that the masses of the 
charged leptons do not depend on the VEV of $\xi$, however 
the mixing angles involved depend on this parameter. 
For $u'_1=u'_2=u' \simeq 10^{-1}$ and $\gamma'$, $\gamma''$ and
 $\beta^{\prime \prime}$ 
of the order unity, we get the charged lepton mixing  angles 
as\footnote{Our choice of 
$u'_1=u'_2 \simeq {\cal O}(10^{-1})$ will be justified  
from  large $y_1$ and $y_2$ in Fig.\ref{fig:paramnh} and 
 large $y_4$ in Fig.\ref{fig:paramih} which we will show
in the next section.}

\be 
\sin \theta^{l}_{12} \simeq \lambda^2,~~~\sin\theta^{l}_{23} 
\simeq 0.1 \lambda,~~~ 
\sin\theta^{l}_{13} \simeq 0.1 \lambda^2  
\label{eq:mixcharged}
~.
\ee
Since $U=U_l^\dagger U_\nu$, where $U$ is the observed lepton mixing 
matrix and $U_\nu$ is the matrix which diagonalizes $m_\nu$ 
given by Eq. (\ref{eq:mnu}), 
the contribution of charged lepton mixing matrix  
would be very tiny. For sin$\theta_{13}$ the 
maximum contribution from the charged lepton is 
${\cal O}(10^{-4})$. In any case, in what follows we show all 
results for $U=U_l^\dagger U_\nu$. 

\section{Phenomenology}

\subsection{Exact ${\mathbf{\mu-\tau}}$ Symmetry Limit}

\begin{figure}[t]
\begin{center}
\includegraphics[height=10.0cm,angle=0]{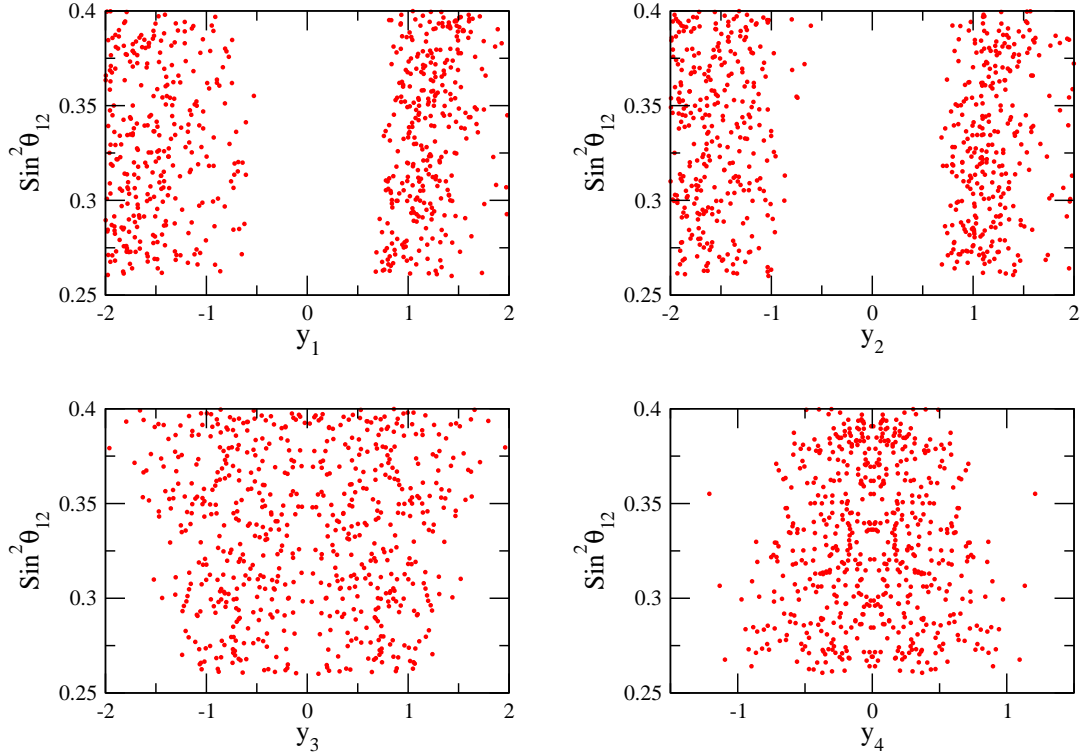}
\caption{\label{fig:sss}
Scatter plots showing the range of the solar mixing angle 
$\sss$ as a function of the model parameters in $m_\nu$ 
in the exact $\mu-\tau$ symmetry limit and for normal hierarchy. 
In each of the panels all other parameters except the one 
appearing in the $x$-axis are allowed to vary freely.   
}
\end{center}
\end{figure}

\begin{figure}[t]
\begin{center}
\includegraphics[height=10.0cm,angle=0]{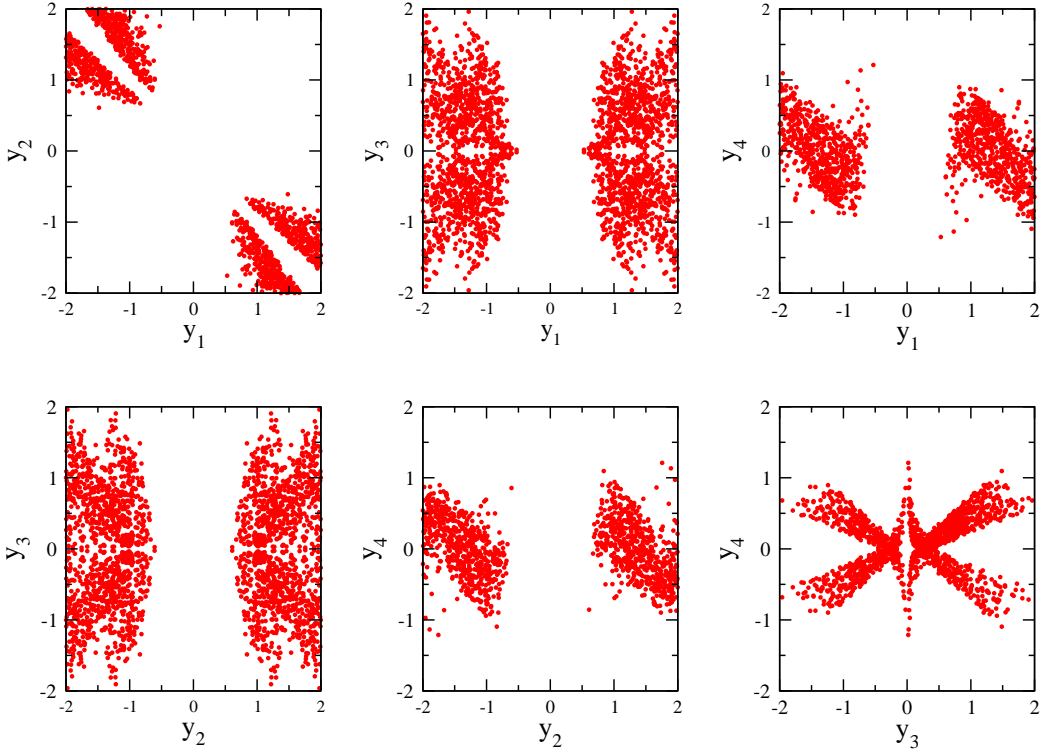}
\caption{\label{fig:paramnh} 
Scatter plots showing allowed ranges of the $m_\nu$ 
model parameters for the normal mass hierarchy 
in the exact $\mu-\tau$ symmetry limit. 
In each of the panels all other parameters except the one 
appearing in the $x$ and y-axes are allowed to vary freely.   
}
\end{center}
\end{figure}

\begin{figure}[t]
\begin{center}
\includegraphics[height=10.0cm,angle=0]{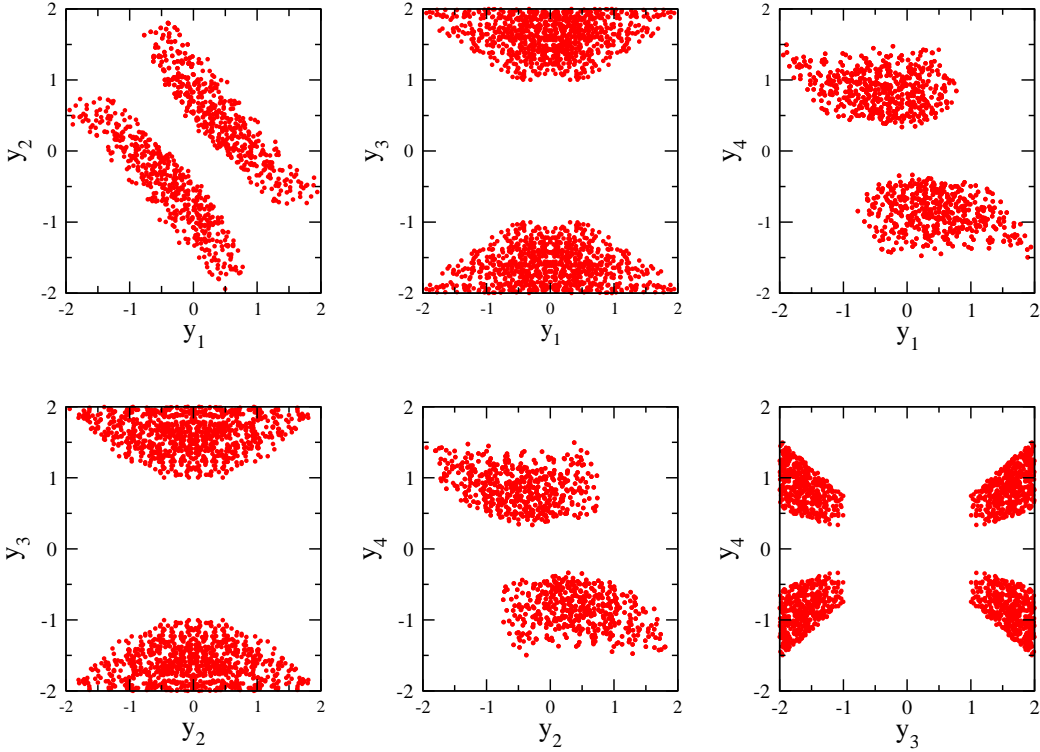}
\caption{\label{fig:paramih}
Same as Figure \ref{fig:paramnh} but for inverted hierarchy.
}
\end{center}
\end{figure}

\begin{figure}[t]
\begin{center}
\includegraphics[height=6.0cm,width=16.0cm,angle=0]{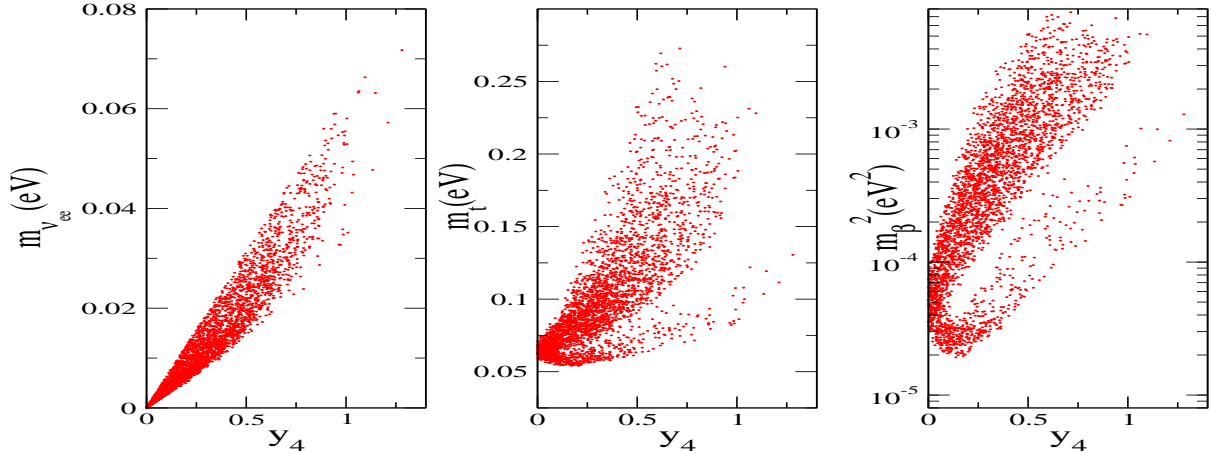}
\caption{\label{fig:mnuee-mtot}
Scatter plots showing variation of $|m_{\nu_{ee}}|$, $m_t$ and 
$m_\beta^2$ with the
model parameter $y_4$.
}
\end{center}
\end{figure}

We have already presented in Eqs. (\ref{eq:massdiff})  
and (\ref{eq:mleigen})
the expressions for the neutrino mass squared differences and  
charged lepton masses, while in Eqs. 
(\ref{eq:mixnu}) and (\ref{eq:mixcharged}) we have given the 
mixing angles in the lepton sector in terms of model parameters. 
We argued that for $\lambda \simeq 2\times 10^{-2} 
\simeq \frac{ \lambda_c^2}{2}$, 
we obtain the charged lepton mass hierarchy in the right ballpark. 

Since neutrino masses are directly proportional 
to $v_1$, it is phenomenologically demanded that the magnitude of 
this VEV should be small. In fact, since $\ma \propto v_1^2$, we take  
$v_1^2 \sim 10^{-4}-10^{-3}$ eV$^2$, and 
find that all experimentally observed neutrino masses 
and mixing constraints are satisfied. It is not unnatural to 
expect such a small value for $v_1^2$. 
For instance, in the most generic 
left-right symmetric models,
\be
v_1 \equiv v_L \sim v_{SM}^2/v_R
~,
\ee
where $v_{SM}$ is the electroweak scale and $v_R$ is the 
VEV of the $SU(2)_R$ Higgs triplet. It is natural to take 
$v_R \sim 10^{13} - 10^{15}$ GeV for which we get 
$v_1 \sim 1 - 10^{-2}$ eV. 

Allowing $v_1^2$ to take any random value between $10^{-4}-10^{-3}$ eV$^2$
we have checked that the neutrino mass spectrum obtained is 
hierarchical. For larger values of $v_1$ of course one 
would get larger values for the absolute neutrino mass scale and 
for $v_1 \sim 1$ eV, we expect a quasi-degenerate neutrino 
mass spectrum. In all our plots we keep $v_1^2$ between 
$10^{-4}-10^{-3}$, eV$^2$. In 
Figure \ref{fig:sss} we show the prediction for $\sss$ as a 
function of the model parameters $y_i$'s. In each panel we 
show the dependence of $\sss$ on a given $y_i$, allowing all 
the others to vary randomly. Here we have assumed normal 
mass hierarchy for the neutrinos. For the charged lepton sector 
we have assumed a fixed set of model parameters which give viable 
charged lepton masses and we took $\lambda \simeq 2\times 10^{-2}$. 
We note that for normal hierarchy ($\ma > 0$): 
\begin{itemize}
\item $y_1=0$ and $y_2=0$ are not allowed.

\item There is almost negligible dependence of $\sss$ on $y_1$ 
and $y_2$ for $|y_1| > 1$ and $|y_2| > 1$ respectively. 

\item The range of $\sss$ decreases with $|y_3|$ and
$|y_4|$. 
\end{itemize}

Figure \ref{fig:paramnh} gives the scatter plots showing allowed 
ranges for the model parameters in two-dimensional parameter 
spaces, taking two parameters at a time and allowing the rest to 
vary freely. We have considered normal hierarchy in this figure.  
We note from the figure that 
for normal hierarchy:
\begin{itemize}
\item $y_1=0$ and $y_2=0$ are not allowed as we had seen before. With 
 $y_1=0$ we would have obtained neutrino mass matrix 
Eq.{(\ref{eq:mnugen})} with two texture zeros in $\mu-\mu$ and $\tau-\tau$
elements  and with $e-\mu$ and $e-\tau$
entries same in the mass matrix  it will not be possible to get a 
normal-hierarchy\cite{tex-mis}. This
could also be explained from the explicit analytical  form of the eigenvalues. 
However, as we will see from  Figure \ref{fig:paramih} inverted hierchy can occur in this case. With $y_2=0$ one gets
 neutrino mass matrix with one texture zero in $\mu-\tau$ element
which is not viable for normal ordering \cite{tex-rodj}.
The allowed values of $y_1$ for normal hierarchy are highly correlated with the 
allowed values of $y_2$ and they are necessarily of opposite signs. 
One obtains a rough linear dependence between the allowed values 
of $y_1$ and $y_2$.


\item $y_4=0$ is allowed and 
there is very little correlation 
of allowed values of $y_4$ with $y_1$ and $y_2$.
For $y_4=0$, one gets a  neutrino mass matrix Eq.{(\ref{eq:mnugen})}
with one texture zero in $e-e$ element which  can produce  normal hierarchy only\cite{tex-rodj}. This predicts $|m_{\nu_{ee}}|=0$.

\item $y_3$ and $y_4$ are strongly correlated. 

\end{itemize}

In Figure \ref{fig:paramih} we show the corresponding allowed ranges 
for the model parameters for  inverted hierarchy. In each of the 
panels, the parameters that do not appear on the $x$ and $y$-axes 
are allowed vary randomly. From a comparison of Figures 
\ref{fig:paramnh} and \ref{fig:paramih} we can observe that the allowed 
areas in the parameter space is almost complementary\footnote{Of course
the same set of model parameter values would {\it never} give 
both normal and inverted hierarchy simultaneously. However, 
we have shown 
two-dimensional projections of the model parameter space and 
hence their could be few overlapping points in the two figures.}.   
We find 
that for the inverted hierarchy:
\begin{itemize}
\item $y_1=0$ and $y_2=0$ {\it simultaneously} are still 
not allowed, though now we can have  $y_1=0$  or  $y_2=0$ 
separately when the other parameter is within a 
certain favorable (non-zero) range. As for normal hierarchy, 
 allowed values of $y_1$ and $y_2$ are highly correlated. As before 
there is a linear dependence between them. 

\item $y_3=0$ and $y_4=0$ are not allowed here. 

In the left panel of Figure  \ref{fig:mnuee-mtot} we show
the variation of the effective neutrino mass $m_{\nu_{ee}}$ with 
the model parameter $y_4$. The effective mass predicted for
neutrino-less double beta decay 
in our model is $|m_{\nu_{ee}}|=|2v_{1}y_{4}|$. 
We have allowed $v_1$ to vary freely in the range  $10^{-1}-10^{-2}$.
From Figure  \ref{fig:mnuee-mtot} one can clearly see that our model predicts
$m_{\nu_{ee}} \lsim 0.07 $ eV. The next generation of 
neutrino-less double experiments are expected to probe 
down to $m_{\nu_{ee}} = 0.01-0.05$ eV \cite{0nbb}.
The middle panel of this figure shows 
the total predicted neutrino mass $m_t$ and right panel shows 
$m_\beta^2$. We find that 
the total neutrino mass $m_t$ (in eV) varies within 
the range $0.05 <m_t<0.28$, while 
the effective mass squared observable 
in beta decay $m_{\beta}^{2} \simeq {\cal O}( 10^{-4}-10^{-2})eV^2$.
The KATRIN experiment will be sensitive to $m_\beta > 0.3$ eV 
\cite{katrin}. 

\end{itemize}

\subsection{Mildly Broken ${\mathbf{\mu-\tau}}$ Symmetry Limit}

\begin{figure}[t]
\begin{center}
\includegraphics[height=15.5cm,angle=270]{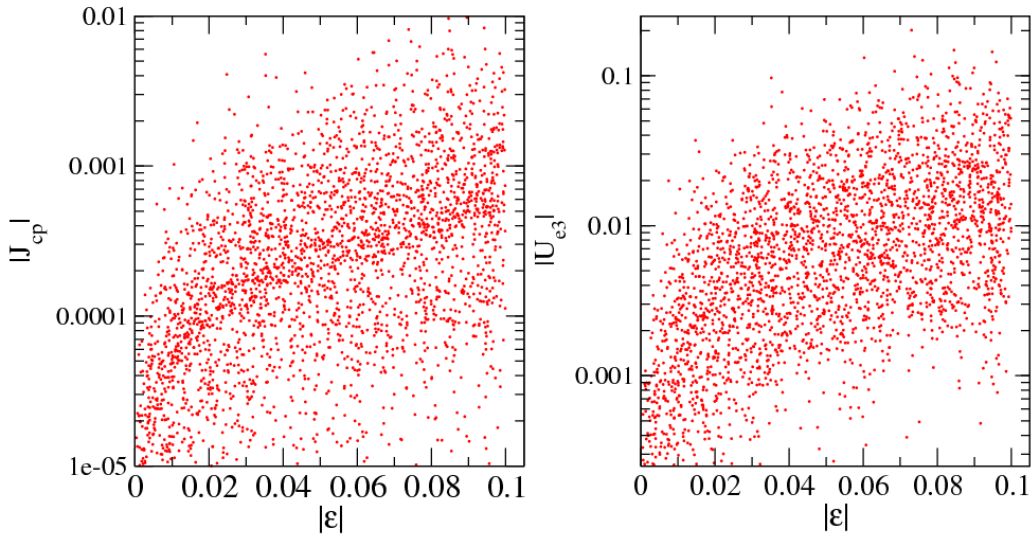}
\caption{\label{fig:jcp}
The Jarlskog invariant $J_{CP}$ (left panel) and 
$\sin\theta_{13}$ as a function of the $\mu-\tau$ 
symmetry breaking parameter $|\epsilon|$. 
}
\end{center}
\end{figure}

So far we have {\it assumed} that the $S_3$ breaking in the 
neutrino sector is such that the residual $\mu-\tau$ 
symmetry is exact. This was motivated by the fact that 
$S_2$ is a subgroup of $S_3$ and we took a particular 
VEV alignment given in Eq. (\ref{eq:nuvev}). 
We will try to justify this 
choice of VEV alignment from minimization of the 
scalar potential. 
In this subsection we will assume that the $\mu-\tau$ 
symmetry is mildly broken. This could come from 
explicit $\mu-\tau$ breaking terms in the Lagrangian. 
In the next section we will see that in our model this 
comes naturally after the minimization of the scalar potential  
due to the deviation of the VEV alignments from 
that given in Eq. (\ref{eq:nuvev}). Small breaking of the 
VEV alignments could also come from radiative corrections and/or higher 
order terms in the scalar  part of the Lagrangian. 
Any breaking of $\mu-\tau$ symmetry will allow $\theta_{23}$ 
to deviate from maximal and $\theta_{13}$ from zero. 
Any non-zero $\theta_{13}$ will open up the possibility of 
low energy CP violation in the lepton sector.  
For the sake of illustration we consider a particular $\mu-\tau$ 
symmetry breaking 
for $m_\nu$ which results from the deviation of the VEV alignment 
from Eq. (\ref{eq:nuvev}). We will see in the next section that 
this deviation is small and could come from $v_1\neq v_2$ 
and/or $u_1\neq u_2$. For the sake of illustration we 
consider only the breaking due to $v_1\neq v_2$. 
We will see that from the minimization of the scalar potential 
one can take $v_1=v_2(1+\epsilon)$. As a result the 
neutrino mass matrix (\ref{eq:numassmatrix}) becomes
\be
m_{\nu}=
2v_{2} \left(
\begin{array}{ccc}
y_{4}u'(2+\epsilon) & y_{3} & y_{3}(1+\epsilon)\\
y_{3} & y_1u' & y_{2}u'(2+\epsilon)\\
y_{3}(1+\epsilon) & y_{2}u'(2+\epsilon) &  y_1u'(1+\epsilon)
\end{array}
\right)
~. 
\ee
We show the values of $|U_{e3}| \equiv \sin\theta_{13}$ 
predicted by the above $m_\nu$ as a function of the 
symmetry breaking parameter $|\epsilon|$ in the right 
panel of Figure \ref{fig:jcp}. 
The left panel panel of this figure shows the 
Jarlskog invariant 
\be
J_{CP} = {\rm Im} \bigg\{U_{e1}\,U_{\mu 2}\,U_{e2}^*\,U_{\mu 1}^* \bigg\}
~,
\ee 
as a function of $|\epsilon|$. We note that the 
model predicts values of $\sin \theta_{13} \lsim 10^{-1}$  
and $J_{CP}\lsim 10^{-2}$ with the exact value determined by the extent of 
symmetry breaking.  This could give 
$\sin^22\theta_{13} \lsim 0.04$, which is just within the 
sensitivity reach of the forthcoming reactor 
experiments like Double Chooz \cite{white} 
and long baseline accelerator 
experiments like T2K \cite{t2k} and NO$\nu$A \cite{nova}. 
These values of $\theta_{13}$ and $J_{CP}$ 
could give a large positive 
signal in the next generation high performance long 
baseline experiments using neutrino beams from 
Neutrino Factories, Superbeams and Beta-beams \cite{issphysics}.

\subsection{Collider Signature and Lepton Flavor Violation}

Recent discussion on the analysis of the scalar potential and the 
Higgs mass specctrum for models with one triplet Higgs 
can be found in \cite{higgstriplet-potential-spectrum}. 
In our model with two Higgs 
triplets we get mixing between the two doubly charged Higgs 
$\Delta_{1}^{++}$ and $\Delta_{2}^{++}$. 
The physical 
Higgs fields can be obtained from the scalar potential 
and are given by
\be
H_{1}^{++}&=&\Delta_{1}^{++}\cos{\theta}\,+\,
\Delta_{2}^{++}\sin{\theta}
~,\\\nonumber
H_{2}^{++}&=&-\Delta_{1}^{++}\sin{\theta}\,+\,
\Delta_{2}^{++}\cos{\theta}
~,
\ee
where the mixing angle $\theta$ is 
\be
\tan2\theta=\frac{e_2(u_1^2+u_2^2) + e_2'u_1u_2}
{h_6'(u_2^2 - u_1^2)
-h_6|v_c|^2}~.
\label{eq:angle}
\ee 
The parameters $e_2$, $e_2'$, $h_6$ 
and $h_6'$ are defined in Eq. (\ref{eq:pot2}).
In the exact $\mu-\tau$ limit, which can be realized by setting
$h_6=0$ and $h_6'=0$ \footnote{We will see this in the next section.}, 
the mixing angle $\theta$ is of course $\pi/4$. 
Even when $h_6$ and $h_6'$ are $\neq 0$, since the couplings 
involved in Eq. (\ref{eq:angle}) are expected to be 
of comparable strengths and since $v_c^2 << u_1^2$ (from the 
observed masses and mixing) and $u_2^2 - u_1^2$ is small 
(see next section) we obtain nearly maximal mixing. 
In the approximate limit where we take $u_1 \simeq  u_2 = u$ and 
neglect $|v_1|^2$, $|v_c|^2$ 
and $v_{SM}^2$ in comparison to $u^2$, 
the square of the masses of the doubly charged Higgs  
are given by
\be
M^2_{H_1^{++}} &\simeq& -a  + u^2\,\left[(4e_1 +2e_1')- (2e_2 + e_2')
\right]
~,
\label{eq:masspp1}
\\
M^2_{H_2^{++}} &\simeq& -a  + u^2\,\left[(4e_1 +2e_1')+ (2e_2 + e_2')
\right]
~,
\label{eq:masspp2}
\ee
The quantities $e_1$, $e_1'$, $e_2$ and $e_2'$ are dimensionless 
coefficients in the scalar potential and will be explained in the 
next section.  
We will see in the next section that $a$ in 
Eqs. (\ref{eq:masspp1}) and  (\ref{eq:masspp2}) 
has a mass dimension 2 and comes as the co-efficient of the 
$\Delta^{++}_{1,2}\Delta^{--}_{1,2}$ term in the scalar potential. We note that 
the masses are modified due the mixing between 
$\Delta^{++}_1$ and $\Delta^{++}_2$, and depend on 
the VEV $u$. The difference between the square of the 
masses of the two doubly charged Higgs depends only on 
$u$ and the coefficients $e_2$ and $e_2'$. Measuring this 
mass squared difference at a collider experiment will provide 
a handle on the VEV $u$, which could then be used in 
conjunction with the lepton mass and mixing data to 
constrain the new scale $\Lambda$.  
In the most natural limit where we take all 
coupling constants $e_1$, $e_1'$, $e_2$ and $e_2'$ to be 
of the same order, then $M^2_{H_1^{++}} \simeq -a $ and 
$M^2_{H_2^{++}} \simeq -a  + 9e_1u^2$.

The most distinctive signature of the existence of triplet Higgs 
can be obtained in collider experiments, through 
the production and subsequent decay of the doubly charged Higgs 
particle(s) \cite{dhdecay,dhdecay1,dhdecay2}. 
The doubly charged Higgs, if produced, would decay 
through the following possible channels:
\be
H^{++} &\rightarrow& H^+H^+ ~,\nonumber\\
H^{++} &\rightarrow& H^+W^+ ~,\nonumber\\
H^{++} &\rightarrow& l^+\,l^+ ~,\nonumber\\
H^{++} &\rightarrow& W^+W^+ ~.
\label{eq:channel}
\ee
Note that our model has two doubly charged and two singly charged Higgs, 
however we have suppressed the corresponding indices in 
Eq. (\ref{eq:channel}). Likewise, we have suppressed the flavor 
indices of the leptons. 
The first two decay modes depend 
on the mass difference between the singly and doubly charged 
Higgs and hence might be kinematically suppressed compared to the 
last two channels. We therefore do not consider them any further. 
The decay rate $H_{1,2}^{++} \rightarrow W^+W^+$ is 
proportional to the square of 
triplet Higgs VEVs $v_{1,2}$, while the 
decay rate to dileptons is inversely proportional to them. 
As a result the ratio of the decay rates for the two channels
is proportional to $v_{1,2}^{-4}$ and  
is given as \cite{dhdecay1}
\be
\frac{\Gamma(H_{1,2}^{++} \rightarrow l_a^+ \l_b^+)}
{\Gamma(H_{1,2}^{++} \rightarrow W^+ W^+)} \approx 
\bigg(\frac{m_\nu}{M_{H_{1,2}^{++}}}\bigg)^2
\bigg(\frac{v_{SM}}{v_{1,2}}\bigg)^4
~,
\ee
where $M_{H_{1,2}^{++}} $ is the mass of
the doubly charged Higgs and $m_\nu$ is the scale of neutrino mass. 
It has been shown \cite{dhdecay1}
from a detailed calculation that for 
for $M_{H_{1,2}^{++}} \simeq 300$ GeV and 
$v_{1,2} \ltap 10^{-4}$ GeV, decay to dileptons will dominate.  
For our model $v_1^2 \approx v_2^2 \simeq 10^{-3} - 10^{-4}$ eV$^2$ and hence we 
can safely neglect decays to $W^+W^+$.   
The decay rate to dileptons is given as \cite{dhdecay,dhdecay1,dhdecay2}
\be
\Gamma(H_{1,2}^{++} \rightarrow l_a^{+} l_b^{+})= 
{\frac{1}{4\pi (1+\delta_{ab})}}|F_{ab}|^{2}M_{H_{1,2}^{++}}
~,
\ee
while the branching ratio for this decay mode is 
\be
BR_{ab}=BR(H_{1,2}^{++} \rightarrow l_a^{+} l_b^{+})=
{\frac{2}{(1+\delta_{ab})}}{\frac{|F_{ab}|^{2}}{\Sigma_{ab}|F_{ab}|^{2}}}
~,
\ee
where $F_{ab}$ are the 
relevant vertex factors which directly depend on the form 
of the neutrino mass matrix. 
Using Eq. (\ref{eq:nulag}), we have tabulated 
in Table \ref{tab:vertices-higgstriplet} the vertex
factors for all possible interaction channels in our model. 
We see that apart from $e$-$\mu$ or 
$e$-$\tau$ combinations given in the table,  
all vertices have extra suppression factor of  
$\frac{u_{1,2}}{\Lambda}$.  
All other vertices arising from Eqs. (\ref{eq:nulag}) and 
(\ref{eq:chlag}) will involve the flavon fields $\xi$ and/or $\phi$ 
and will be suppressed by higher orders in 
$\Lambda$. We therefore do not give them here.
We had argued from Eq. (\ref{eq:angle}) that in 
the exact $\mu-\tau$ symmetric limit $\theta=\pi/4$. 
One can then immediately see from 
Table \ref{tab:vertices-higgstriplet}  
that $H_2^{++}ee$ and $H_2^{++} \mu\tau$ couplings are zero. 
Therefore, in the exact $\mu-\tau$ symmetric limit, the decay of 
$H_2^{++}$ to $ee$ and $\mu\tau$ is strictly forbidden. 
We have noted above that even when we do not 
impose exact $\mu-\tau$ symmetry, $\theta\approx \pi/4$
and hence these decay channels will be suppressed. 
The branching ratio of all the other decay modes 
are determined by the corresponding Yukawa couplings. 
Generally speaking, 
since all the vertices other than  
$H_{1,2}^{++}e\mu$ and $H_{1,2}^{++}e\tau$ are $\frac{u_{1,2}}{\Lambda}$
suppressed, branching ratio of these channels will be larger than 
all others, assuming equal values of $y_1$,$y_2$,$y_3$ and $y_4$.
However,  $y_3$ and $y_4$ could be small for 
normal hierarchy while inverted hierarchy could be 
produced for very small $y_1$ and $y_2$. 
This will give a handle on 
determining the neutrino parameters in general and 
the neutrino mass hierarchy in particular \cite{dhdecay2}. 
For instance, 
if the decay modes of doubly charged Higgs to $e\mu$ and $e\tau$ 
are not observed at a collider experiment, then it would imply 
small $y_3$, which would disfavor the inverted hierarchy. 

Signature of doubly charged Higgs could in principle also be seen in 
lepton flavor 
violating processes. 
However, in the framework of our model 
all additional contribution to 
$l_i \rightarrow l_j\gamma$ 
are smaller than what is expected in the standard model. One 
can check from Table \ref{tab:vertices-higgstriplet} that 
the only additional diagram which does not have any $\Lambda$ 
(or $u_{1,2}/\Lambda$) suppression contributes to 
$\tau \rightarrow \mu\gamma$. However, even this diagram will 
be suppressed due to $M_{H_{1,2}^{++}} \gg M_W$ \cite{radiativedecay}. 
The presence of $H_{1,2}^{++}$ will allow the decay modes 
of the form $l_l \rightarrow l_il_jl_k$ at the tree level,
where $l_i$, $l_j$ and $l_k$ are leptons of any flavor. 
The branching ratios for $\mu\rightarrow eee$ 
and $\tau\rightarrow eee$ in our model for exact $\mu-\tau$ 
symmetry is given by \cite{mutoeee}
\be
BR(\mu \rightarrow eee) \simeq \frac{1}{16\,G_F^2}\frac{u_1^2}{\Lambda^2}
\frac{|y_4^*y_3|^2}{M_{H_{1}^{++}}^4}
~.
\ee
Thus we see that even this process is suppressed by $u_1^2/\Lambda^2$ 
compared to other models with triplet Higgs. Branching ratio for 
all other lepton flavor violating decay modes such as 
$\tau \rightarrow \mu\mu\mu$ are further suppressed. 
The only decay mode which comes unsuppressed is 
$\tau \rightarrow ee\mu$, for which the 
branching ratio is given by
\be
BR(\tau \rightarrow ee\mu) \simeq \frac{1}{4\,G_F^2}
\frac{|y_3|^4}{M_{H_{1,2}^{++}}^4}
~.
\ee
The current experimental constraint on this decay mode is 
$BR(\tau \rightarrow ee\mu) < 2 \times 10^{-7}$ \cite{pdg}, 
which constrains our model parameter $y_3$ as (assuming 
$M_{H_{1,2}^{++}} \sim 300$ GeV)
\be
|y_3| \ltap 10^{-1}
~.
\label{eq:boundlfv}
\ee
Recall that $y_3$ 
is predicted to be large for the inverted hierarchy while it 
could be tiny for the normal hierarchy. On the face of it 
then it appears that the bound given by Eq. (\ref{eq:boundlfv}) 
disfavors the inverted hierarchy for our model. However, 
recall that the allowed values of $y_3$ shown in Figs. 
\ref{fig:paramnh} and \ref{fig:paramih} were presented 
assuming $v_1^2$ to lie between $10^{-3}-10^{-4}$ eV$^2$. 
However, $v_1^2$ could be higher and since what determines the 
mass squared differences $\ms$ and $\ma$ is the product of 
$v_1^2$ and the Yukawas, higher $v_1^2$ would imply smaller values 
of the latter. For instance, we could have taken $v_1^2 \sim 10^{-1}-10^{-2}$
eV$^2$ and in that case inverted hierarchy would still be allowed.
We reiterate that the bound given by Eq. (\ref{eq:boundlfv}) has 
been obtained assuming $M_{H_{1,2}^{++}}\sim 300$ GeV. For 
more massive doubly charged Higgs the braching ratio would 
go down.  
On the other hand, if one uses bounds from lepton flavor violating 
decays to constrain the Yukawas, then one would obtain 
corresponding limits on the value of $v_1$. 
We conclude that with improved bounds on lepton flavor violating 
decay modes, one could test our model and/or the neutrino mass hierarchy 
predicted by our model.

\begin{table} 
\centering
\begin{tabular}{|c|c|c|}
\hline
Vertices & Vertex factors $F_{ab}$  \\
\hline
$e\mu$ $H_1^{++}$ & $2y_3sin\theta CP_L$  \\
\hline
$e\mu$$ H_2^{++}$ & $2y_3cos\theta CP_L$  \\
\hline
$e\tau$$ H_1^{++}$ & $2y_3cos\theta CP_L$\\
\hline
$e\tau H_2^{++} $& $2y_3sin\theta CP_L$\\
\hline
$eeH_1^{++} $&$ y_4 \frac{(sin\theta u_1+cos\theta u_2)}{\Lambda}CP_L$\\
\hline

$eeH_2^{++} $&$ y_4 \frac{(cos\theta u_1-sin\theta u_2)}{\Lambda}CP_L$\\
\hline
$\mu \tau H_1^{++} $&$ y_2 \frac{(sin\theta u_1+cos\theta u_2)}{\Lambda}CP_L$\\
\hline
$\mu \tau H_2^{++} $&$ y_2 \frac{(cos\theta u_1-sin\theta u_2)}{\Lambda}CP_L$\\
\hline
$\tau \tau H_1^{++} $&$ y_1\frac{u_1}{\Lambda}cos\theta CP_L$\\
\hline
$\tau \tau H_2^{++} $&$ y_1\frac{u_1}{\Lambda}sin\theta CP_L$\\
\hline
$\mu \mu H_1^{++}$ & $y_1\frac{u_2}{\Lambda}sin\theta CP_L$\\
\hline
$\mu \mu H_2^{++} $&$ y_1\frac{u_2}{\Lambda}cos\theta CP_L$\\
\hline
\end{tabular}
\caption{Doubly charged 
Higgs triplet and lepton vertices and the corresponding 
vertex factors $F_{ab}$, where a and b are generation indices. 
The  charged lepton mass matrix is almost diagonal in our model. 
In this analysis we have considered that mass basis and flavor basis of the charged
leptons are the same.}
\label{tab:vertices-higgstriplet}
\end{table}

\section{The Vacuum Expectation Values}
\label{sec:scalar}

Up to terms of dimension four, the $S_3\times Z_4\times Z_3$ 
invariant scalar potential (cf. Table \ref{tab:charges3}) is given by 
\be
V= \sum_{i} V_{i}
\label{eq:pot1}
\ee
where 
\be
V_{1} &=& -a Tr[\Delta' \Delta] +b(Tr[\Delta' \Delta])^{2}  \nonumber    \\
V_{2}^a&=&[-c(\xi \xi)+h.c]+c^{\prime}(\xi^{\prime}\xi) \nonumber \\
V_{2}^b&=&[d(\xi \xi)^{\underline 1}(\xi \xi)^{\underline1}+h.c]+d^{\prime}(\xi^{\prime} \xi^{\prime})^{\underline 2}(\xi \xi)^{\underline 2}+[d^{\prime \prime}(\xi^{\prime} \xi)^{\underline 1}(\xi \xi)^{\underline1}+h.c]  \nonumber \\
V_{3}^{a}&=&[e_{1}Tr[(\Delta' \Delta)^{\underline 1}] 
(\xi \xi)^{\underline1}+h.c]+e_{1}^{\prime}Tr[(\Delta' \Delta)^{\underline 1}] 
(\xi^{\prime} \xi)^{\underline1} \nonumber \\
V_{3}^{b}&=&[e_{2}Tr[(\Delta' \Delta)^{\underline 2}] 
(\xi \xi)^{\underline 2} + h.c]+e_{2}^{\prime}Tr[(\Delta' \Delta)^{\underline 2}] 
(\xi^{\prime} \xi)^{\underline 2} \nonumber  \\
V_{3}^{c}&=& h_{6}^{\prime}
Tr[\Delta' \Delta]^{\underline 1'}(\xi^{\prime} \xi)^{\underline 1'}+ h_{6}^{\prime \prime}(\xi^{\prime} \xi)^{\underline 1'}(\phi'_{e} \phi_{e})^{\underline 1'}\nonumber \\
V_{4}&=&f_{1}Tr[(\Delta' \Delta)^{\underline 1}
(\Delta' \Delta)^{\underline 1}]+f_{2}Tr[(\Delta' \Delta)^{\underline 1'}
(\Delta' \Delta)^{\underline 1'}]+f_{3}Tr[(\Delta' \Delta)^{\underline 2}
(\Delta' \Delta)^{\underline 2}] \nonumber    \\
V_{5}&=&-h_{1}(\phi'_{e} \phi_{e})+ h_{2}(\phi'_{e} \phi'_{e})^{\underline 1}
(\phi_{e} \phi_{e})^{\underline 1}+h_{3}
(\phi'_{e} \phi'_{e})^{\underline 2}(\phi_{e} \phi_{e})^{\underline 2} 
\nonumber  \\
V_{6}&=& h_{4} Tr[\Delta' \Delta]^{\underline 1}
(\phi'_{e} \phi_{e})^{\underline 1} + h_{5}Tr[\Delta' \Delta]^{\underline 2}
(\phi'_{e} \phi_{e})^{\underline 2} +   h_{6}
Tr[\Delta' \Delta]^{\underline 1'}(\phi'_{e} \phi_{e})^{\underline 1'}      
\nonumber \\
V_{7}^{a}&=&[l_{1} (\xi \xi)^{\underline 1}(\phi'_{e} \phi_{e})^{\underline 1}+h.c] 
+l_{1}^{\prime \prime} (\xi^{\prime} \xi)^{\underline 1}(\phi'_{e} \phi_{e})^{\underline 1} \nonumber \\
V_{7}^{b}&=&[l_{2}(\xi \xi)^{\underline 2}(\phi'_{e} \phi_{e})^{\underline 2} +h.c]
+l_{2}^{\prime} (\xi^{\prime} \xi)^{\underline 2}(\phi'_{e} \phi_{e})^{\underline 2}+l_{4}(H^\dagger H)(\phi'_{e}\phi_{e}) 
\nonumber \\
V_{8}&=&a_{1}Tr[\Delta' \Delta](H^\dagger
H)+[a_{2}(H^\dagger H)(\xi
\xi)+h.c]-\mu^2(H^\dagger H)+\lambda(H^\dagger H)^2 \nonumber \\
 &&+r(H^{\dagger}\tau_{i}H)Tr[\Delta^{\prime}\tau_{i}\Delta]+a_{2}^{\prime \prime}(H^\dagger H)(\xi^{\prime}\xi)
\label{eq:pot2} 
\ee
The underline sign in the superscript represents the 
particular $S_3$ representation from the tensor product of 
the two $S_{3}$ doublets. The superscripts ``2'' without 
the underline represent the square of the term. The quantities 
with primes are obtained following Eq. (\ref{eq:primes})
\be
\xi' = \sigma_{1}(\xi)^{\dagger}=\left(\begin{array}{c}
\xi^{\dagger}_{2}\\
\xi^{\dagger}_{1}
\end{array}
\right)
~,~~~
\phi'_{e} = \sigma_{1}(\phi_{e})^{\dagger}=\left(\begin{array}{c}
\phi^{\dagger}_{2}\\
\phi^{\dagger}_{1}
\end{array}
\right)
~,~~~
\Delta' = \sigma_{1}(\Delta)^{\dagger}=\left(\begin{array}{c}
\Delta^{\dagger}_{2}\\
\Delta^{\dagger}_{1}
\end{array}
\right)
~.
\ee
The potential given by Eqs. (\ref{eq:pot1}) and (\ref{eq:pot2}) 
has to be minimized.  
 The singlets  $\xi$ and $\phi_e$ pick up VEVs which 
spontaneously breaks the $S_3$ symmetry at some high scale,
while $\Delta$ picks up a VEV when 
$SU(2)_L\times U(1)_Y$ is broken at the electroweak scale. 
The VEVs have already been given 
in Eqs. (\ref{eq:xivev}), (\ref{eq:deltavev}), and (\ref{eq:phivev}). 
Though we begin by taking the VEV alignment for 
$\langle \phi_e \rangle$ in order to obtain the 
VEV alignments for $\langle \Delta \rangle$ and $\langle \xi \rangle$, 
we stress that this does not pose any serious threat to our 
model and it can be shown to be perfectly self-consistent. 
For the sake of keeping the algebra simple 
we take the VEVs of $\Delta$ and $\phi_e$  
to be complex but the VEVs of $\xi$ to be real. We have 
checked that a complex VEV for $\xi$ does not bring 
any qualitative change to our calculations.

We denote $ v_{1}= |v_{1}|e^{i\alpha_{1}}, 
v_{2}= |v_{2}|e^{i\alpha_{2}}$, where $v_1$ and $v_2$ are the 
VEVs of $\Delta_1$ and $\Delta_2$. 
Substituting this in 
Eqs. (\ref{eq:pot1}) and (\ref{eq:pot2}) we obtain 
\be
V&=&(-a+4e_{1}u_{1}u_{2}+e_1^{\prime}(u_1^2+u_2^2)+h_{4}|v_c|^{2}+a_{1}v_{SM}^{2})
(|v_{2}|^{2}+|v_{1}|^{2})+(b+f_{1}+f_{2})
(|v_{2}|^{2}+|v_{1}|^{2})^{2}\nonumber\\
&&-4cu_{1}u_{2}+c^{\prime}(u_1^2+u_2^2)+8du^{2}_{1}u^{2}_{2}+d^{\prime}(u_1^4+u_2^4)+4d^{\prime \prime}u_1u_2(u_1^2+u_2^2)+2(f_{3}-2f_{2})|v_{1}|^{2}|v_{2}|^{2}\nonumber \\
&&+2|v_{1}||v_{2}|[e_2(u^{2}_{1}+u^{2}_{2})+e_2^{\prime}u_1u_2]\,
\cos(\alpha_{2}- \alpha_{1}) +(-h_{1}+4l_{1}u_{1}u_{2})|v_c|^{2}+l_1^{\prime \prime}|v_c|^{2}(u_1^2+u_2^2)\nonumber \\
&&+h_{3}|v_c|^4+4a_{2}v_{SM}^{2}u_{1}u_{2}+[-h_{6}|v_c|^{2}+h_6^{\prime}(u_{2}^{2}-u_{1}^{2})](|v_{2}|^{2}-|v_{1}|^{2})-h_6^{\prime \prime}(u_{2}^{2}-u_{1}^{2})|v_c|^{2}\nonumber \\
&&+a_2^{\prime \prime}v_{SM}^{2}(u_1^2+u_2^2)   +l_{4}
v_{SM}^{2}|v_c|^{2}-\mu^2v_{SM}^{2}+\lambda v_{SM}^{4}
\ee
where we have absorbed $r$ in the redefine $a_1$.
The minimization conditions are:

\be
\frac{\partial V}{\partial (\alpha_{2} - \alpha_{1})} = 0
~,
\label{eq:cond1}
\ee 
\be
\frac{\partial V}{\partial (|v_{1}|)} = 0
~,
\label{eq:cond2}
\ee 
\be
\frac{\partial V}{\partial (|v_{2}|)} = 0
~,
\label{eq:cond3}
\ee 
\be
\frac{\partial V}{\partial u_{1}} = 0
~,
\label{eq:cond4}
\ee 
\be
\frac{\partial V}{\partial u_{2}} = 0 
~,
\label{eq:cond5}
\ee
\be
\frac{\partial V}{\partial |v_c|} = 0
~.
\label{eq:cond6}
\ee
From Eq. (\ref{eq:cond1}) we obtain the condition, 
\be
2|v_{1}||v_{2}|[e_2(u^{2}_{1}+u^{2}_{2})+e_2^{\prime}u_1u_2]\,\sin(\alpha_{2}-\alpha_{1})=0
~.
\ee
Hence 
\be
\alpha_{2} = \alpha_{1}
~,
\label{eq:vevalignarg}
\ee  
as long as $ |v_{1}| , |v_{2}|$ and 
$[e_{2}(u^{2}_{1}+u^{2}_{2})+e_2^{\prime}u_1u_2]$ are $ \neq 0$.
Eq. (\ref{eq:cond2}) leads to the condition, 
\be
&&-2a|v_{1}|+4B(|v_{2}|^{2}+|v_{1}|^{2})|v_{1}|+2|v_{1}|[4e_{1}u_{1}u_{2}+e_1^{\prime}(u_1^{2}+u_2^{2})]+2|v_{2}|[e_{2}(u_{2}^{2}+u_{1}^{2})+e_{2}^{\prime}u_1u_2]\nonumber \\
&&+4F|v_{1}||v_{2}|^{2}+2h_4|v_{1}||v_c|^{2}+2a_{1}|v_{1}|v_{SM}^{2}+2h_6|v_c|^{2}|v_{1}|-2h_{6}^{\prime}|v_1|(u_{2}^{2}-u_{1}^{2})=0
~,
\label{eq:c2}
\ee
where we have defined $ B=(b+f_{1}+f_{2})$,  $F=f_{3}-2f_{2}$ 
and we have used $\alpha_{2}= \alpha_{1}$. 
Using Eq. (\ref{eq:cond3}) we obtain, 
\be
&&-2a|v_{2}|+4B(|v_{2}|^{2}+|v_{1}|^{2})|v_{2}|+2|v_{2}|[4e_{1}u_{1}u_{2}+e_1^{\prime}(u_1^{2}+u_2^{2})]+2|v_{1}|[e_{2}(u_{2}^{2}+u_{1}^{2})+e_{2}^{\prime}u_1u_2]\nonumber \\      
&&+4F|v_{2}||v_{1}|^{2}+2h_4|v_{2}||v_c|^{2}+2a_{1}|v_{2}|v_{SM}^{2}-2h_6|v_c|^{2}|v_{2}|+2h_{6}^{\prime}|v_1|(u_{2}^{2}-u_{1}^{2})=0
~.
\label{eq:c3}
\ee
Multiplying Eq. (\ref{eq:c2}) by $|v_{2}|$ and 
Eq. (\ref{eq:c3}) by   $|v_{1}|$ and 
subtracting one from the other we obtain,
\be
 (2e_{2}(u^{2}_{1}+u^{2}_{2})+2e_2^{\prime}u_1u_2+4F|v_{1}||v_{2}|)(|v_{1}|^{2}-|v_{2}|^{2})=4|v_{1}||v_{2}|[h_{6}|v_c|^{2}-h_6^{\prime}(u_{2}^{2}-u_{1}^{2})]
~.
\label{eq:vevalign}
\ee
In the limit $h_6 = 0$  and $h_6^{\prime}=0$ we get $|v_{2}|=|v_{1}|$ 
(if $e_{2}$, $e_2^{\prime}$ and $F \neq 0$ simultaneously), 
which is required for exact $\mu$-$\tau$ symmetry in 
the neutrino sector. However, there is no {\it a priori}
reason to assume that $h_6$, and $h_6^{\prime}$ are zero. 
In the most general case 
keeping non-zero  $h_6$ and $h_6^{\prime}$, we obtain
\be
|v_1|^2=|v_2|^2+ \frac{4|v_{1}||v_{2}|[h_6|v_c|^{2}-h_6^{\prime}(u_2^{2}-u_{1}^{2})]}
{2e_2(u_{1}^2+u_{2}^2)+2e_2^{\prime}u_1u_2+4F|v_{1}||v_{2}|}
~.
\ee
Since $|v_{1}||v_{2}| \ll u_1u_2$ and $(u_{1}^2+u_{2}^2)$ we neglect the 
$4F|v_{1}||v_{2}|$ term from the denominator. 
For $u_1\simeq u_2=u$ and $e_2\simeq e_2^{\prime}$, one obtains 
\be
|v_1|^2=|v_2|^2+ \frac{2|v_{1}||v_{2}|h_6|v_c|^{2}}
{3e_2u^{2}}
~.
\ee
For a fixed $v_2$, this is a quadratic equation in $v_1$  which 
allows the solution $v_1\simeq v_2(1+\epsilon)$ where 
$\epsilon=\frac{h_6v_c^{2}}{3e_2u^2}$. For $h_6$ and $e_2$ 
of the same order and $\frac{u}{\Lambda}=10^{-1}$,
$\frac{v_c}{\Lambda}=10^{-2}$ we obtain $\epsilon \simeq 10^{-2}\ll 1$ .
This would give rise to a very mild breaking of the $\mu-\tau$ symmetry. 
We have discussed this case in section 3.2. 

Using Eqs. (\ref{eq:cond4}) and (\ref{eq:cond5}) 
and repeating the same excercise we get the deviation from $u_1=u_2$ as
\be
u_{1}^{2}=u_{2}^{2}+\frac{A}{B}
\label{eq:c4a}
\ee
where $A$ and $B$ are 
\be
A=4u_1u_2[h_6^{\prime \prime}|v_c|^{2}-h_6^{\prime}(|v_2|^{2}-|v_1|^{2})]
\label{eq:c5a}
\ee
and 
\be
B &=&(-4c+16du_1u_2-4d^{\prime}u_1u_2+4d^{\prime \prime}(u_1^{2}+u_2^{2})+4e_{1}(|v_{1}|^{2}+|v_{2}|^{2})+2e_2^{\prime}|v_1||v_2| +4l_{1}|v_c|^{2} \nonumber \\
&&+4a_{2}v_{SM}^2)
\label{eq:c5b}
~,
\ee
and using the same arguments as above, it is not hard to see that the 
deviation from $u_1 = u_2$ is also mild. Again, $u_1=u_2$ is 
satisfied when $h_6'=0$ and $h_6^{\prime\prime}=0$. 
Since $h_6=0$ is also required for $|v_1|=|v_2|$ to be satisfied, 
we conclude that exact $\mu-\tau$ symmetry for neutrinos demands that 
$h_6=0$,  $h_6'=0$ and $h_6^{\prime\prime}=0$ simultaneously. 
\\

\noindent
Finally, 
from the last minimization condition (\ref{eq:cond6}) we get the solution,
\be 
|v_c|^{2}= \frac{1}{4h_{3}}&
\bigg[&\!\!2h_{6}(|v_{2}|^{2}-|v_{1}|^{2})+2h_{6}^{\prime
    \prime}(u_{2}^{2}-u_{1}^{2})-2l_1^{\prime
    \prime}(u_1^{2}+u_2^{2})+2h_{1}
\nonumber\\
&&-2h_{4}(|v_{1}|^{2}+|v_{2}|^{2})-8u_{1}u_{2}l_{1}-2l_{4}v_{SM}^2\bigg]
~.
\label{eq:c7}
\ee 

We next use use the condition (\ref{eq:c7}) to estimate the 
cut-off scale $\Lambda$. Since 
$h_1$ define in Eq. (\ref{eq:pot2}) 
gives the square of the mass of the $\phi_e$ fields,  
it could be large. 
The other couplings $h_3$, $l_1$, $l_4$, $h_4$, $h_6^{\prime \prime}$, $l_1^{\prime \prime}$ and $h_6$ 
are dimensionless and  
can be assumed to have roughly the same order of magnitude 
which should be much much  smaller than $h_1$. 
Dividing both sides of Eq. (\ref{eq:c7}) by
$\Lambda^2$ and using $|v_1|^2 \simeq |v_2|^2 = 10^{-3}$ eV$^2$, 
$\frac{v_c}{\Lambda}\simeq 2\times 10^{-2}$,
$ \frac{u_{1,2}}{\Lambda}\simeq 10^{-1}$ and hence
${(\frac{v_1}{\Lambda})}^2 \ll {(\frac{ v_c}{\Lambda})}^2 
< {(\frac {u_1}{\Lambda})}^2$ , we get 
\be
\Lambda^2 \simeq \frac {h_1}{4l_1+2l_1^{\prime \prime}}\times 10^{2}
~{\rm GeV}^2
~.
\ee
Note that $h_1$ has mass dimension 2 and in principle could be large.
If we take $h_1$ in TeV range, for example if we take $\sqrt{h_1}=10$ 
TeV, then the cut-off scale of the theory is fixed as
$10^2$ TeV, where we have taken $l_1$ and $l_1^{''} \simeq {\cal O}(1)$.
From $\frac{u_1}{\Lambda}=\frac{u_2}{\Lambda}\sim 10^{-1}$
and $\frac{v_c}{\Lambda}\sim 10^{-2}$, we then obtain $u_{1,2}=10$ TeV 
and $v_c=1$ TeV. We reiterate that the constraints from the lepton 
masses themselves do not impose any restriction on the cut-off 
scale and the VEVs. One can obtain estimates on them only through 
limits on the masses of the Higges. For instance, from Eqs. 
(\ref{eq:masspp1}) and  (\ref{eq:masspp2}) one could 
in principle estimate 
$u$ by measuring the difference between doubly charged Higgs masses. 
This could then be combined with the neutrino data to get $\Lambda$, and 
finally use the charged lepton masses to get $v_c$. 

Since we consider a model with triplet Higgs to generate Majorana 
neutrino masses, it is pertinent to make 
some comments regarding breaking of lepton number 
and possible creation of a 
massless goldstone called Majoron \cite{majoron}. 
If we wish to conserve lepton number in our effective 
Lagrangian giving neutrino masses (cf. Eq. (\ref{eq:nulag})), 
we would have to assign lepton number $-2$ to our Higgs fields 
$\Delta$, and lepton number 0 to the flavor fields $\xi$. 
One could then fear that Majorons would be created when 
$\Delta$ get VEVs. However, lepton number is only an accidental symmetry 
of the standard model. It is possible that this symmetry was 
broken in the theory at the high scale. 
In any case, we do not consider lepton number to be a good 
symmetry of our theory. For instance, 
we could break it explicitly 
by giving a lepton number to the fields $\xi$. In that case 
one would not break lepton number spontaneously and there would be 
no Majoron.

These VEV alignments have been obtained by assuming no effect 
of renormalization group running. However, it is understood 
that the running from the high scale where $S_3$ is broken 
to the electroweak scale where the masses are generated, 
will modify the VEV alignments. Another way the VEV alignments 
could get modified is through higher dimensional terms in the 
scalar potential. Due to the $Z_4$ as well as $Z_3$ symmetry that 
we have imposed, one cannot get terms of dimension five in the 
scalar potential. The possible next order terms in $V$ would 
therefore be terms of dimension six. These terms would be suppressed 
by $\Lambda^2$ and are therefore expected to be much less important in $V$.

\section{Conclusions}

We have attempted to provide a viable model for the lepton masses and 
mixing by imposing a $S_3\times Z_4 \times Z_3$ family symmetry. 
Our model requires two $SU(2)_L$ Higgs triplets 
arranged in the doublet representation of $S_3$. In addition we 
need 2 sets of flavon $S_3$ doublets which are singlets with respect 
to the standard model. By suitably arranging our fermions 
in the different representations of $S_3$, we constructed the 
Yukawa part of the neutrino and charged lepton Lagrangian. 
Desired structure for the mass matrices were obtained by 
giving suitable $Z_4$ and $Z_3$ charges to the particles. 
In particular, the most common dimension five operator 
$LLHH$, 
\footnote{Here $L$ represents the lepton doublet and $H$ the standard 
model Higgs doublet.}, which gives Majorana 
neutrino masses is strictly forbidden in our 
model by the flavor symmetry. 
This term arises in seesaw models.
Type-I seesaw is forbidden by $Z_4$ symmetry and since it 
would have required right-handed 
standard model singlet 
neutrinos (at a high scale), 
these are therefore naturally absent in our model.
Type-II seesaw would require the coupling 
$\Delta HH$ which is 
forbidden by the $S_3$ flavor symmetry.   
Due to the presence of the Higgs triplets, we have Majorana neutrino 
masses from dimension four operator $l_1D_l\Delta$ itself. In addition, 
we have masses generated by other dimension five operators and together 
they provide a phenomenological correct form for the neutrino mass matrix. 
The observed charged lepton masses can be obtained 
very naturally from our model.  

Neutrino data demands $\theta_{23}$ to be maximal and $\theta_{13}$ 
to be zero. This hints towards the presence of $\mu-\tau$ 
symmetry in the neutrino sector. On the other hand the 
wide disparity between the $\mu$ and $\tau$ masses demands 
that this exchange symmetry does not exists for the 
charged leptons. 
The $\mu-\tau$ 
reflection symmetry is a subgroup of $S_3$ and 
we break $S_3$ in such a way that this exchange symmetry is 
preserved for the neutrinos, while it is maximally violated for 
the charged lepton. This allowed for simultaneous explanation of 
the peculiar 
mixing pattern of the neutrinos and the strong hierarchical 
mass pattern for the charged lepton. One requires a certain 
alignment for the VEVs for the Higgses for achieving this. 
The VEV alignment in our model 
comes  from the minimum condition of the 
scalar potential without any need for imposing supersymmetry 
and/or additional driving fields. We performed an explicit minimization 
of the scalar potential to justify the VEV alignment required. 
We showed that exact $\mu-\tau$ symmetry could be obtained 
if certain conditions are satisfied. In the most general case, 
we obtained a deviation from exact $\mu-\tau$ symmetry. However, 
we showed that these deviations are very extremely small in our model. 

We studied the phenomenological viability and predictions of our 
model in the exact $\mu-\tau$ symmetric limit and produced 
plots showing correlations between the different model parameters. 
We gave predictions for $\sss$, $\ms$, $\ma$, effective mass 
in neutrino-less double beta decay, the observed mass squared in 
direct beta decay and total mass of the neutrinos relevant to 
cosmological data. 
We also allowed for mild breaking of the $\mu-\tau$ symmetry and 
calculated $\theta_{13}$ and the strength of CP violation in the 
lepton sector. We showed results for one illustrative symmetry 
breaking scenario and concluded that one would be 
able to observe such small $\theta_{13}$ and CP violation in
 next-generation 
long baseline experiments involving powerful beams. 

Our model predicts lepton flavor violating 
processes such at $\tau \rightarrow ee\mu$ at the 
tree level. This and other lepton flavor violating  
processes could therefore be used to constrain 
the model as well as the neutrino mass hierarchy. 
Production and subsequent decay of the doubly charged Higgs 
at particle colliders is a smoking gun signal for the 
existence of triplet Higgs. We showed that in our model 
since the triplet VEV is required to be very small, 
decay to dileptons would predominate. The lepton flavors 
involved in the final lepton pair could be used to distinguish 
this model from the other models with triplet Higgs. 
These signatures could also be used to distinguish 
the inverted and normal hierarchy.

\vglue 0.8cm
\noindent
{\Large{\bf Acknowledgments}}\vglue 0.3cm
\noindent
The authors are extremely grateful to A. Raychaudhuri for discussions.  
M.M. wishes to thank A. Sen, Y. Lin and A. Watanabe for discussions.We specially thank Christoph Luhn for a very useful correspondence. 
The authors acknowledge the HRI cluster facilities for computation.
This work has been supported by the Neutrino Project 
under the XI Plan of Harish-Chandra Research Institute. 

\section{Appendix: The $S_3$ Permutation Symmetry Group}

\begin{table} 
\centering
\begin{tabular}{|c|c|ccc|}
\hline\hline
& & & &   \cr
Conjugacy Class & Elements & $~~~1~~~$& $~~~1^\prime~~~$ & $~~~2~~~$ \cr
& & & & \cr
\hline\hline
& & & &   \cr
$C_1$  & $e$ & 1 & 1 & 2 \cr
& & & &   \cr
$C_2$ & $(1\,2),(2\,3),(1\,3)$ & 1 & -1 & 0 \cr
& & & &  \cr
$C_3$ & $(1\,2\,3),(3\,2\,1)$ & 1 & 1 & -1 \cr
& & & &   \cr
\hline
\end{tabular}
\caption{\label{tab:class} 
Character table of $S_3$. The first column gives the 
classes, the second gives the elements in each class, 
and last three 
columns give the character corresponding to the three 
irreducible representations $1$, $1^\prime$ and 2. 
 }
\end{table} 
The group $S_3$ is the permutation group of three distinct objects, 
and is the smallest non-abelian symmetry group. It consists of a 
set of rotations which leave an equilateral triangle 
invariant in three dimensions. The group has six elements 
divided into three conjugacy classes. 
The generators of the group are $S$ and $T$ which 
satisfy
\be
S^2 = T^3 = (ST)^2 = 1
~.
\ee
The elements are given by the permutations 
\be
G \equiv \bigg\{e,(1\,2),(1\,3),(2\,3),(1\,2\,3),(3\,2\,1)\bigg\}
~,
\ee
which can be written in terms of the generators as
\be
G \equiv \bigg\{e,ST,S,TS,T^2,T\bigg\}
~.
\ee
One can see that the $S_3$ group contains two kinds of subgroups. 
It can be easily checked that the subgroup of elements
\be
G_{Z_3} \equiv \bigg\{e,T,T^2\bigg\}
~,
\ee
form a group under $Z_3$. In addition, there are three 
$S_2$ permutation subgroups\footnote{The group $S_2$ is 
isomorphic to $Z_2$.} 
\be
G_{S_{12}} \equiv \bigg\{e,(1\,2)\bigg\}~,~~~~
G_{S_{13}} \equiv \bigg\{e,(1\,3)\bigg\}~,~~~~
G_{S_{23}} \equiv \bigg\{e,(2\,3)\bigg\}
~.
\ee
In this paper we are mainly interested in the 
permutation subgroup 
which corresponds to 
$\mu-\tau$ exchange symmetry. 
We will break the $S_3$ group for neutrinos in such a way that 
$\mu-\tau$ symmetry 
remains intact once neutrino mass terms are generated after 
electroweak symmetry breaking. 
On the other hand, for the charged leptons we will 
break it maximally in order to generate the desired hierarchy 
between the $\mu$ and $\tau$ masses. 

The group contains two one-dimensional and one two-dimensional 
irreducible representations. 
The one-dimensional 
representations are given by
\be
1: ~~~&S=1,&~~T=1
\\
1^\prime:~~~&S=-1,&~~T=1
~~.
\ee
The two-dimensional representation is given by
\be
2:~~~S=\pmatrix{0&1\cr 1&0},~~~~T=\pmatrix{\omega & 0 \cr 
0 & \omega^2}
~.
\ee
%
The character table is given in Table \ref{tab:class}. 
Using the Table we can write down the rules for the tensor products. 
For the one-dimensional irreducible representations we have 
\be
1\times 1 = 1,~~~~
1\times 1^\prime = 1^\prime,~~~~
1^\prime \times 1^\prime = 1
~.
\ee
Tensor products between two doublets $\psi = (\psi_1,\psi_2)^T$ 
and $\phi = (\phi_1,\phi_2)^T$ are given as
\be
2\times 2 = 1 + 1^\prime + 2
~,
\label{eq:product}
\ee
where
\be
1&~\equiv~&\psi_1\phi_2 + \psi_2\phi_1~,
\\
1^\prime&~\equiv~&\psi_1\phi_2 - \psi_2\phi_1~,
\\
2&~\equiv~&\pmatrix{\psi_2\phi_2 \cr  \psi_1\phi_1}
~.
\ee
The complex conjugate doublet $\psi^\star$ is given as $2^\star$ 
for which the generators are $S^\star$ and $T^\star$. One can 
easily check that $\psi^\star$ does not transform as doublet (2) of 
$S_3$ and therefore 
for this case a meaningful way of writing the tensor 
products for the conjugate fields is by defining 
\be
\psi^\prime \equiv \sigma_1 \psi^\star = \pmatrix{\psi_2^\star 
\cr \psi_1^\star}
~.
\label{eq:primes}
\ee
Using the relations $\sigma_1S^\star\sigma_1 = S$ 
and $\sigma_1T^\star\sigma_1 = T$ 
one can show that $\psi^\prime$ transforms as a doublet. Then 
the tensor products $\psi^\prime \times \phi$ 
are given by Eq. (\ref{eq:product}) where
\be
1&~\equiv~&\psi_1^\star\phi_1 + \psi_2^\star\phi_2~,
\\
1^\prime&~\equiv~&\psi_2^\star\phi_2 - \psi_1^\star\phi_1 ~,
\\
2&~\equiv~&\pmatrix{\psi_1^\star\phi_2 \cr  \psi_2^\star\phi_1}
~.
\ee



\begin{thebibliography}{99}

\bibitem{review}
See for example, 
  M.~C.~Gonzalez-Garcia and M.~Maltoni,
  Phys.\ Rept.\  {\bf 460}, 1 (2008)
  arXiv:0704.1800 [hep-ph].

\bibitem{limits}
  A.~Bandyopadhyay, S.~Choubey, S.~Goswami, S.~T.~Petcov and D.~P.~Roy,
  arXiv:0804.4857 [hep-ph].
%
M.~Maltoni, 
T.~Schwetz, M.~A.~Tortola and J.~W.~F.~Valle,
New J.\ Phys.\  {\bf 6}, 122 (2004), hep-ph/0405172 v6;
%
  G.~L.~Fogli {\it et al.},
  arXiv:0805.2517 [hep-ph].

\bibitem{mutau}
 T.~Fukuyama and H.~Nishiura, hep-ph/9702253; 
R.~N.~Mohapatra and S.~Nussinov, 
Phys.\ Rev.\ D {\bf 60}, 013002 (1999); 
E.~Ma and M.~Raidal, Phys. Rev. Lett. {\bf 87}, 011802 (2001); 
C.~S.~Lam, Phys.\ Lett.\ B {\bf 507}, 214 (2001); 
P.F. Harrison and W. G. Scott, 
Phys.\ Lett.\ B {\bf 547}, 219 (2002); 
T.~Kitabayashi and M.~Yasue, 
Phys.\ Rev.\ D {\bf 67}, 015006 (2003); 
W.~Grimus and L.~Lavoura, 
Phys.\ Lett.\ B {\bf 572}, 189 (2003); 
J.\ Phys.\ G {\bf 30}, 73 (2004); 
Y.~Koide, 
Phys.\ Rev.\ D {\bf 69}, 093001 (2004); 
A. Ghosal, hep-ph/0304090; 
W.~Grimus {\it et al.}, 
Nucl.\ Phys.\ B {\bf 713}, 151 (2005); 
R.~N.~Mohapatra, 
JHEP {\bf 0410}, 027 (2004); 
A.~de Gouvea, Phys.\ Rev.\ D {\bf 69}, 093007 (2004); 
 R.~N.~Mohapatra and W.~Rodejohann,
  Phys.\ Rev.\ D {\bf 72}, 053001 (2005); 
 R.~N.~Mohapatra and S.~Nasri, 
  Phys.\ Rev.\ D {\bf 71}, 033001 (2005); 
R.~N.~Mohapatra, S.~Nasri and H.~B.~Yu, 
Phys.\ Lett.\ B {\bf 615}, 231 (2005); 
  Phys.\ Rev.\ D {\bf 72}, 033007 (2005); 
Y.~H.~Ahn{\it et al.}, 
  Phys.\ Rev.\ D {\bf 73}, 093005 (2006); 
  B.~Brahmachari and S.~Choubey,
  Phys.\ Lett.\  B {\bf 642}, 495 (2006);
  K. Fuki, M. Yasue, 
  hep-ph/0608042.

\bibitem{lmultau}
  S.~Choubey and W.~Rodejohann,
  Eur.\ Phys.\ J.\  C {\bf 40}, 259 (2005);
%
  B.~Adhikary,
  Phys.\ Rev.\  D {\bf 74}, 033002 (2006);
%
T.~Ota and W.~Rodejohann,
  Phys.\ Lett.\ B {\bf 639}, 322 (2006).



\bibitem{tbm} 
  P.~F.~Harrison, D.~H.~Perkins and W.~G.~Scott,
  Phys.\ Lett.\  B {\bf 530}, 167 (2002);
%
  P.~F.~Harrison, D.~H.~Perkins and W.~G.~Scott,
  Phys.\ Lett.\  B {\bf 458}, 79 (1999);
%
P.~F.~Harrison and W.~G.~Scott,
arXiv:hep-ph/0402006.
%
Z.~z.~Xing,
Phys.\ Lett.\ B {\bf 533} (2002) 85;
%
  X.~G.~He and A.~Zee,
  Phys.\ Lett.\  B {\bf 560}, 87 (2003);
%
  L.~Wolfenstein,
  Phys.\ Rev.\  D {\bf 18}, 958 (1978).

\bibitem{a4} 
  E.~Ma and G.~Rajasekaran,
  Phys.\ Rev.\  D {\bf 64}, 113012 (2001);
%
  E.~Ma,
  Mod.\ Phys.\ Lett.\  A {\bf 17}, 627 (2002);
%
  K.~S.~Babu, E.~Ma and J.~W.~F.~Valle,
  Phys.\ Lett.\  B {\bf 552}, 207 (2003);
%
K. S. Babu and X.-G. He, 
  arXiv:hep-ph/0507217;
  E.~Ma,
  Mod.\ Phys.\ Lett.\  A {\bf 22}, 101 (2007);
%
  E.~Ma,
  Mod.\ Phys.\ Lett.\  A {\bf 21}, 2931 (2006);
%
  E.~Ma, H.~Sawanaka and M.~Tanimoto,
  Phys.\ Lett.\  B {\bf 641}, 301 (2006);
%
  B.~Adhikary, B.~Brahmachari, A.~Ghosal, E.~Ma and M.~K.~Parida,
  Phys.\ Lett.\  B {\bf 638}, 345 (2006);
%
  E.~Ma,
  Phys.\ Rev.\  D {\bf 73}, 057304 (2006);
%
  S.~L.~Chen, M.~Frigerio and E.~Ma,
  Nucl.\ Phys.\  B {\bf 724}, 423 (2005);
%
  E.~Ma,
  Phys.\ Rev.\  D {\bf 72}, 037301 (2005);
%
  E.~Ma,
  Mod.\ Phys.\ Lett.\  A {\bf 20}, 2601 (2005);
%
  E.~Ma,
  Phys.\ Rev.\  D {\bf 70}, 031901 (2004);
%
  L.~Lavoura and H.~Kuhbock,
  Mod.\ Phys.\ Lett.\  A {\bf 22}, 181 (2007);
%
  Y.~Koide,
  Eur.\ Phys.\ J.\  C {\bf 52}, 617 (2007);
%
  S.~F.~King and M.~Malinsky,
  Phys.\ Lett.\  B {\bf 645}, 351 (2007);
%
  X.~G.~He, Y.~Y.~Keum and R.~R.~Volkas,
  JHEP {\bf 0604}, 039 (2006);
%
  A.~Zee,
  Phys.\ Lett.\  B {\bf 630}, 58 (2005);
%
  F.~Bazzocchi, S.~Kaneko and S.~Morisi,
  arXiv:0707.3032 [hep-ph].
%
  B.~Adhikary and A.~Ghosal,
  Phys.\ Rev.\  D {\bf 75}, 073020 (2007);
%
  M.~Hirsch, J.~C.~Romao, S.~Skadhauge, J.~W.~F.~Valle and A.~Villanova del Moral,
  Phys.\ Rev.\  D {\bf 69}, 093006 (2004);
%
  M.~Hirsch, A.~S.~Joshipura, S.~Kaneko and J.~W.~F.~Valle,
  Phys.\ Rev.\ Lett.\  {\bf 99}, 151802 (2007);
%
  M.~Honda and M.~Tanimoto,
  arXiv:0801.0181 [hep-ph].

\bibitem{AF}
  G.~Altarelli and F.~Feruglio,
  Nucl.\ Phys.\  B {\bf 720}, 64 (2005);
%
  G.~Altarelli and F.~Feruglio,
  Nucl.\ Phys.\  B {\bf 741}, 215 (2006);
 G.~Altarelli , F.~Feruglio and Y.~Lin,
  Nucl.\ Phys.\  B {\bf 775}, 31 (2007).
%


\bibitem{a4us}
  B.~Brahmachari, S.~Choubey and M.~Mitra,
  Phys.\ Rev.\  D {\bf 77}, 073008 (2008). 

\bibitem{a4extension}
  G.~Altarelli, F.~Feruglio and C.~Hagedorn,
  JHEP {\bf 0803}, 052 (2008);
%
  F.~Bazzocchi, S.~Morisi, M.~Picariello and E.~Torrente-Lujan,
  arXiv:0802.1693 [hep-ph];
%
  S.~Morisi, M.~Picariello and E.~Torrente-Lujan,
  Phys.\ Rev.\  D {\bf 75}, 075015 (2007);
%
  L.~Lavoura and H.~Kuhbock,
  Eur.\ Phys.\ J.\  C {\bf 55}, 303 (2008);
%
  E.~Ma, H.~Sawanaka and M.~Tanimoto,
  Phys.\ Lett.\  B {\bf 641}, 301 (2006);
%
  X.~G.~He, Y.~Y.~Keum and R.~R.~Volkas,
  JHEP {\bf 0604}, 039 (2006);
%
  M.~C.~Chen and K.~T.~Mahanthappa,
  Phys.\ Lett.\  B {\bf 652}, 34 (2007);
%
Y.~Lin,
arXiv:0804.2867[hep-ph].

\bibitem{s3old}
  E.~Ma,
  Phys.\ Rev.\  D {\bf 44}, R587 (1991);
%
  Y.~Koide,
  Phys.\ Rev.\  D {\bf 60}, 077301 (1999);
%
  M.~Tanimoto,
  Phys.\ Lett.\  B {\bf 483}, 417 (2000);
%
  J.~Kubo,
  Phys.\ Lett.\  B {\bf 578}, 156 (2004)
  [Erratum-ibid.\  B {\bf 619}, 387 (2005)];
%
  F.~Caravaglios and S.~Morisi,
  arXiv:hep-ph/0503234;
%
  S.~Morisi and M.~Picariello,
  Int.\ J.\ Theor.\ Phys.\  {\bf 45}, 1267 (2006).
%

\bibitem{s3hs}
  P.~F.~Harrison and W.~G.~Scott,
  Phys.\ Lett.\  B {\bf 557}, 76 (2003).


\bibitem{s3z2}
  W.~Grimus and L.~Lavoura,
  JHEP {\bf 0508}, 013 (2005).

\bibitem{s3tbm}
  R.~N.~Mohapatra, S.~Nasri and H.~B.~Yu,
  Phys.\ Lett.\  B {\bf 639}, 318 (2006).

\bibitem{s3DMT}
 N.~Haba and K.~Yoshioka ,
 Nucl.\ Phys.\ B {\bf 739},254 (2006).


\bibitem{s3}
  C.~Y.~Chen and L.~Wolfenstein,
  Phys.\ Rev.\  D {\bf 77}, 093009 (2008);
%
  S.~Kaneko, H.~Sawanaka, T.~Shingai, M.~Tanimoto and K.~Yoshioka,
  arXiv:hep-ph/0703250;
%
  Y.~Koide,
  Eur.\ Phys.\ J.\  C {\bf 50}, 809 (2007);
%
  Y.~Koide,
  Phys.\ Rev.\  D {\bf 73}, 057901 (2006);
%
  T.~Teshima,
  Phys.\ Rev.\  D {\bf 73}, 045019 (2006);
%
  F.~Caravaglios and S.~Morisi,
  arXiv:hep-ph/0503234;
%
  L.~Lavoura and E.~Ma,
  Mod.\ Phys.\ Lett.\  A {\bf 20}, 1217 (2005);
%
  T.~Araki, J.~Kubo and E.~A.~Paschos,
  Eur.\ Phys.\ J.\  C {\bf 45}, 465 (2006);
%
N.~Haba, A.~Watanabe and K.~Yoshioka ,
Phys.\ Rev.\ Lett {\bf 97}, 041601 (2006).



\bibitem{s3FL}
  F.~Feruglio and Y.~Lin,
  Nucl.\ Phys.\  B {\bf 800}, 77 (2008)
  [arXiv:0712.1528 [hep-ph]].

\bibitem{anomaly}
C.~Luhn and P.~Ramond ,
 arXiv:0805.1736 [hep-ph].

\bibitem{lelmlt}
S.~T.~Petcov,
Phys.\ Lett.\ B {\bf 110}, 245 (1982); for more recent studies  
see, e.g., 
R.~Barbieri \textit{et al.}, 
JHEP {\bf 9812}, 017 (1998); 
A.~S.~Joshipura and S.~D.~Rindani, 
Eur.\ Phys.\ J.\ C {\bf 14}, 85 (2000); 
R.~N.~Mohapatra, A.~Perez-Lorenzana and C.~A.~de Sousa Pires, 
Phys.\ Lett.\ B {\bf 474}, 355 (2000); 
Q.~Shafi and Z.~Tavartkiladze, 
Phys.\ Lett.\ B {\bf 482}, 145 (2000).
L.~Lavoura,
Phys.\ Rev.\ D {\bf 62}, 093011 (2000); 
W.~Grimus and L.~Lavoura,
Phys.\ Rev.\ D {\bf 62}, 093012 (2000); 
T.~Kitabayashi and M.~Yasue, 
Phys.\ Rev.\ D {\bf 63}, 095002 (2001); 
A.~Aranda, C.~D.~Carone and P.~Meade,
Phys.\ Rev.\ D {\bf 65}, 013011 (2001); 
K.~S.~Babu and R.~N.~Mohapatra, 
Phys.\ Lett.\ B {\bf 532}, 77 (2002); 
H.~J.~He, D.~A.~Dicus and J.~N.~Ng, 
Phys.\ Lett.\ B {\bf 536}, 83 (2002)
H.~S.~Goh, R.~N.~Mohapatra and S.~P.~Ng, 
Phys.\ Lett.\ B {\bf 542}, 116 (2002); 
G.~K.~Leontaris, J.~Rizos and A.~Psallidas, 
Phys.\ Lett.\ B {\bf 597}, 182 (2004). 

\bibitem{zeeus}
  B.~Brahmachari and S.~Choubey,
  Phys.\ Lett.\  B {\bf 642}, 495 (2006).

\bibitem{tex-mis}

  P.~H.~Frampton, S.~L.~Glashow and D.~Marfatia,
  Phys.\ Lett.\  B {\bf 536}, 79 (2002);
  Z.~Z.~Xing,
  Phys.\ Lett.\  B {\bf 530}, 159 (2002);
  B.~Desai, D.~P.~Roy and A.~R.~Vaucher,
  Mod.\ Phys.\ Lett. \ A {\bf 18}, 1355 (2003).
  

\bibitem{tex-rodj}
  A.~Merle and W.~Rodejohanm,
  Phys.\ Rev.\  D {\bf 73}, 073012 (2006).


\bibitem{0nbb}
  F.~T.~Avignone,
  Nucl.\ Phys.\ Proc.\ Suppl.\  {\bf 143}, 233 (2005).

\bibitem{katrin}
 A.~Osipowicz {\it et al.}  [KATRIN Collaboration],
  arXiv:hep-ex/0109033.

\bibitem{white}
  K.~Anderson {\it et al.},
  arXiv:hep-ex/0402041.

\bibitem{t2k}
  Y.~Itow {\it et al.},
  arXiv:hep-ex/0106019.

\bibitem{nova}
  D.~S.~Ayres {\it et al.}  [NOvA Collaboration],
  arXiv:hep-ex/0503053.

\bibitem{issphysics}
  A.~Bandyopadhyay {\it et al.}  [ISS Physics Working Group],
  arXiv:0710.4947 [hep-ph].

\bibitem{higgstriplet-potential-spectrum}
  P.~Dey, A.~Kundu and B.~Mukhopadhyaya,
  arXiv:0802.2510 [hep-ph].
 
\bibitem{dhdecay}
  A.~Abada{\it et al},
  JHEP 0712:061 (2007);
  A.~G.~Akeroyd, M.~Aoki and H.~Sugiyama,
  Phys.\ Rev. \ D {\bf77}, 075010 (2008). 

\bibitem{dhdecay1}
  P.~ Fileviez{\it et al.}
  Phys.\ Rev.\  D {\bf 78}, 015018 (2008).

\bibitem{dhdecay2}
  J.~Garayoa and T.~Schwetz,
  JHEP 0803:009 (2008).


\bibitem{radiativedecay}
  S.~M.~Bilenky and S.~T.~Petcov,
  Rev.\ Mod.\ Phys.\  {\bf 59}, 671 (1987)
  [Erratum-ibid.\  {\bf 61}, 169.1989\ ERRAT,60,575 (1989\ ERRAT,60,575-575.1988)];
  R.~N.~Mohapatra,
  Phys.\ Rev. \ D {\bf46}, 2990 (1992). 

\bibitem{mutoeee}
  M.~Kakizaki, Y.~Ogura and F.~Shima,
  Phys.\ Lett.\  B {\bf 566}, 210 (2003).

\bibitem{pdg}
  W.~M.~Yao {\it et al.}  [Particle Data Group],
  J.\ Phys.\ G {\bf 33} (2006) 1.

\bibitem{majoron}
  G.~B.~Gelmini and M.~Roncadelli,
  Phys.\ Lett.\  B {\bf 99}, 411 (1981);
  Y.~Chikashige, R.~N.~Mohapatra and R.~D.~Peccei,
  Phys.\ Lett.\  B {\bf 98}, 265 (1981);
  J.~Schechter and J.~W.~F.~Valle,
  Phys.\ Rev. \ D {\bf22}, 2227 (1980);
  T.~P.~Cheng and L.~F.~Li,
  Phys.\ Rev. \ D {\bf22}, 2860 (1980).



\end{thebibliography}
\end{document}